\documentclass[%
 reprint,
superscriptaddress,
 amsmath,amssymb,
 aps,
]{revtex4-2}

\usepackage{graphicx}% Include figure files
\usepackage{dcolumn}% Align table columns on decimal point
\usepackage{bm}% bold math
\usepackage{hyperref}% add hypertext capabilities
\begin{document}

\preprint{APS/123-QED}

\title{Electron energy loss and angular asymmetry induced by elastic scattering in helium droplets}% Force line breaks with \\

\author{Jakob D. Asmussen}
\affiliation{Department of Physics and Astronomy,
	Aarhus University, Denmark}
\author{Keshav Sishodia}
\affiliation{Department of Physics,
	Indian Institute of Technology Madras, India}
\author{Björn Bastian}
\affiliation{Department of Physics and Astronomy,
	Aarhus University, Denmark}
 \author{Abdul R. Abid}
\affiliation{Department of Physics and Astronomy,
	Aarhus University, Denmark}
\author{Ltaief Ben Ltaief}
\affiliation{Department of Physics and Astronomy,
	Aarhus University, Denmark}
 \author{Henrik B. Pedersen}
\affiliation{Department of Physics and Astronomy,
	Aarhus University, Denmark}
\author{Subhendu De}
\affiliation{Department of Physics,
	Indian Institute of Technology Madras, India}
\author{Christian Medina}
\affiliation{Institute of Physics, 
	University of Freiburg, Germany}
\author{Nitish Pal}
\affiliation{Elettra Sincrotrone, Trieste, Italy}
\author{Robert Richter}
\affiliation{Elettra Sincrotrone, Trieste, Italy}
 \author{Thomas Fennel}
 \affiliation{Institut für Physik, 
 Universität Rostock, Germany}
\author{Sivarama Krishnan}
\affiliation{Department of Physics,
	Indian Institute of Technology Madras, India}
\author{Marcel Mudrich}
\email{mudrich@phys.au.dk}
\affiliation{Department of Physics and Astronomy,
	Aarhus University, Denmark}

%\affiliation{%

% \collaboration{MUSO Collaboration}%\noaffiliation

%  \homepage{http://www.Second.institution.edu/~Charlie.Author}
% \affiliation{
%  Second institution and/or address\\
%  This line break forced% with \\
% }%
% \affiliation{
%  Third institution, the second for Charlie Author
% \affiliation{%
%  Authors' institution and/or address\\
%  This line break forced with \textbackslash\textbackslash
% }%

% \collaboration{CLEO Collaboration}%\noaffiliation

\date{\today}% It is always \today, today,

\begin{abstract}
Helium nanodroplets are ideal model systems to unravel the complex interaction of condensed matter with ionizing radiation. Here we study the effect of purely elastic electron scattering on angular and energy distributions of photoelectrons emitted from He nanodroplets of variable size ($10$-$10^9$ atoms per droplets). For large droplets, photoelectrons develop a pronounced anisotropy along the incident light beam due to a shadowing effect within the droplets. In contrast, the detected photoelectron spectra are only weakly perturbed. This opens up possibilities for photoelectron spectroscopy of dopants embedded in droplets provided they are smaller than the penetration depth of the light and the trapping range of emitted electrons.
\end{abstract}

\maketitle

Irradiation of condensed matter with ionizing radiation induces various secondary processes in addition to the primary photoionization event. In biological matter, secondary processes such as electron-impact ionization, radical reactions and dissociative attachment of low-energy electrons (LEEs) to biomolecules are the main causes for radiation damage leading to mutations or cell death~\cite{gomez2012radiation,alizadeh2015biomolecular}. Deciphering the mechanisms of radiation damage is crucial for devising improved schemes for radiotherapy for cancer~\cite{hanel2003electron,martin2004dna,gokhberg2014site,Stumpf:2016}. 
LEEs are mainly produced by scattering of the primary and secondary electrons in the medium~\cite{bass1998absolute}.
% and to some extent by energy-transfer processes~\cite{Jahnke:2020}. 
In bulk molecular systems, complex interactions of electrons with intra- and intermolecular degrees of freedom make it hard to unravel the various elastic and inelastic scattering channels. Furthermore, the tendency of slow electrons to remain trapped in the material hinders their detection. Therefore, solid or liquid nanoparticles (NPs) have been used to study the dynamics of electron scattering~\cite{ban2020photoemission} and other properties making NPs relevant for atmospheric, astrophysical, and technical sciences~\cite{niessner1989laboratory,ziemann1998secondary,wilson2006vuv,wilson2007size,signorell2016nanofocusing}.
%Photoelectron emission of NPs has also been studied to determine ionization energies, for aerosol analysis, and to probe molecular coatings on aerosol particles in the context of atmospheric, astrophysical, and technical sciences where NPs play important roles~\cite{niessner1989laboratory,ziemann1998secondary,wilson2006vuv,wilson2007size,signorell2016nanofocusing}.

As atoms or molecules condense into clusters and NPs, photoelectron angular distributions (PADs) tend to become more isotropic because emitted electrons scatter on the cluster constituents and thereby change their emission direction. A reduction of the detected electron kinetic energy for increasing NP size is typically ascribed to inelastic scattering;
energy transfer is small in elastic electron-molecule collisions due to the large mass mismatch ($<1/1836$). Nevertheless, elastic scattering can lead to diffusion-like, damped electron motion in extended media, as we show in this work. At particle sizes $\gtrsim100$~nm, electron distributions are additionally altered by optical confinement effects such as nanofocusing and shadowing~\cite{signorell2016nanofocusing}. The latter occurs when the particle size exceeds both the penetration depth of the radiation and the mean path traveled by an emitted electron before it is trapped in the droplet (`trapping range'). Electrons are then preferentially emitted from that side of the particle facing toward the incident light~\cite{watson1973photoelectron,wilson2007size}. In electron imaging experiments, the degree of shadowing is often specified by the anisotropy parameter $\alpha = I_{f}/I_{b}\leq 1$, where $I_{f/b}$ is the electron intensity in the forward/backward half plane of the electron image with respect to the incoming light. So far, the various phenomena related to electron scattering have been studied separately for different types of NPs~\cite{signorell2016nanofocusing}. The continuous evolution of photoelectron distributions from individual atoms up to large NPs has not been reported. In particular, the effect of elastic scattering on electron distributions has not been  demonstrated explicitly.

In this work, we present a comprehensive study of photoemission spectra (PES) and PAD of large superfluid helium nanodroplets (HNDs). HNDs are particularly well suited model systems to study electron scattering as they feature an extraordinarily wide gap in their excitation spectrum reaching from $\sim1$~eV (phonons) up to 20~eV (excitons)~\cite{toennies2004superfluid}. This gives us the unique opportunity to investigate purely elastic electron-atom scattering in a condensed-phase system in a wide energy range. Additional complications due to nano-focusing effects~\cite{signorell2016nanofocusing} are negligible as the refractive index of He only weakly deviates from 1 by $<0.02$~\cite{henke1993x,chantler1995theoretical}. Due to their superfluid nature, HNDs are homogeneous, spherically symmetric particles featuring a flat, nearly size-independent density distribution~\cite{harms1998density,gomez2014shapes}. The droplet size can be varied continuously in a wide range from a few nm up to $\gtrsim10~\mu $m~\cite{gomez2011sizes,kolatzki2022micrometer}. 
HNDs are widely used as cold, inert and transparent cryo-matrices to form tailored molecular complexes, metal clusters and NPs which can be probed by spectroscopy and mass spectrometry~\cite{toennies2004superfluid,slenczka2022molecules}. Extending the HND technique to extreme-ultraviolet or X-ray photoelectron spectroscopy (UPS, XPS) of embedded (`dopant') species is an intriguing prospect, in particular for probing novel types of nano-aggregates formed in HNDs~\cite{boatwright2013helium,haberfehlner2015formation,schiffmann2020helium,messner2020shell,slenczka2022molecules}. However, it has remained an open question to what extent PES from dopants are distorted by the interaction of emitted electrons with the He matrix. Previous PES measurements of dopants using ns and fs multi-photon laser-ionization schemes appeared to be largely unperturbed by the surrounding He~\cite{radcliffe2004excited,loginov2005photoelectron,kazak2019photoelectron,ltaief2020direct,ltaief2021photoelectron,dozmorov2018quantum,thaler2018femtosecond}, while Penning ionization electron spectra are notoriously broad and structure-less~\cite{wang2008photoelectron,shcherbinin2018penning,mandal2020penning,ltaief2021photoelectron}. Likewise, recent experiments investigating above-threshold ionization of atoms and molecules in HNDs indicated that electron-He scattering substantially impacts PES and PAD~\cite{treiber2021observation,michiels2021enhancement,krebs2022phase,treiber2022dynamics,zhou2023enhancing}. The relaxation of hot electrons at energies $>1$~eV in HNDs has been described by electron-atom binary collisions~\cite{loginov2005photoelectron}. Below $1$~eV, electrons localize in void bubbles~\cite{onn1969attenuation,onn1971injection,sethumadhavan2004detection} which then scatter at elementary modes of the HND (phonons and rotons) before emerging to the droplet surface where electrons are eventually released into the vacuum~\cite{sethumadhavan2004detection,aitken2017thermodynamic}.
%In a pioneering work on photoelectron spectroscopy out of HNDs by Loginov et al.~\cite{loginov2005photoelectron}, the authors conjectured that the electron energy loss was mainly due to elastic binary collisions; However, due to the small range of droplet sizes and the fixed electron energy accessible in that experiment, a definitive conclusion on the nature of the electron relaxation was impossible. Here, we study both photoelectron spectra and electron angular distributions in a wide range of He droplet sizes and photon energies. 
Here we show experimentally and using classical scattering simulations that electron-He elastic scattering is an important mechanism leading to substantial electron-energy loss (EEL) and drastically altered PADs of emitted electrons. Large HNDs with radii $R\gtrsim20$~nm containing $\langle N\rangle\gtrsim10^6$ He atoms feature a pronounced anisotropy of the electron PAD along the light propagation axis, but the PES of detected electrons are only weakly perturbed. %Electron-energy loss up to about 15\,\%.

\begin{figure}[h]
    \centering
    \includegraphics[width=0.95\columnwidth]{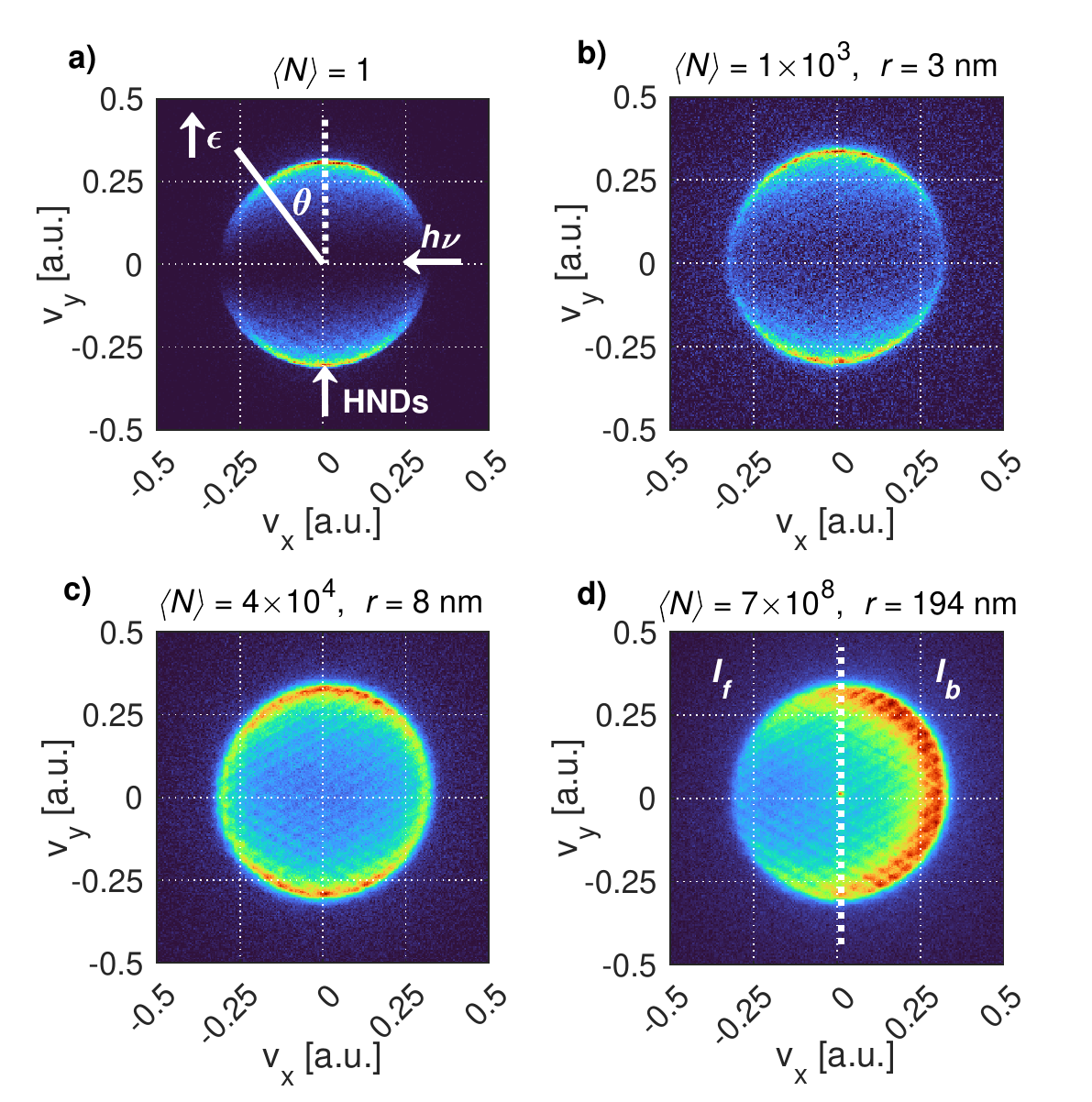}
    \caption{a) VMIs of photoelectrons detected in coincidence with He$^+$ from He gas. Arrows show the light propagation direction ($h\nu$), polarization ($\epsilon$) and the droplet beam direction (HNDs). $\theta$ is the angle of photoemission with respect to the polarization of the light. b)-d) VMIs of all electrons emitted from a beam of HNDs of various sizes. All VMIs are measured at $h\nu=26$~eV. In d) the dotted line separates the forward ($I_f$) and backward ($I_b$) half-planes of the image with respect to the incident light. The grid pattern visible in c) and d) is a detector artifact.   %\textbf{e)} shows the absorption of the incoming light in a cross-section of a droplet of size $\overline{r} = 170$~nm.   
    }
    \label{fig:VMIs}
\end{figure}

HNDs were formed by continuous expansion of He at high pressure (30-50~bar) into vacuum through a cryogenically cooled (7-32~K) nozzle of diameter $5~\mu$m. By controlling the expansion pressure and temperature, the droplet mean radius $R$ was varied in the range 1-190~nm ($\langle N \rangle = 10$-$10^9$). The droplet size was determined by titration measurements~\cite{gomez2011sizes} and by comparison to literature values~\cite{toennies2004superfluid}. 
The XENIA (XUV electron-ion spectrometer for nanodroplets in Aarhus) endstation~\cite{bastian2022new} located at the AMOLine of the ASTRID2 synchrotron at Aarhus University, Denmark~\cite{hertel2011astrid2}, was used to record velocity-map images (VMIs) of photoelectrons. PES and PAD were inferred from the VMIs by Abel inversion using the MEVELER reconstruction method~\cite{dick2014inverting}. Additionally, high resolution PES were measured using a hemispherical analyzer (VG-220i) mounted under the magic angle at the GasPhase beamline at the synchrotron Elettra in Trieste, Italy~\cite{blyth1999high}.
%\subsection{Monte-carlo simulations} 

%The He number density inside the spherical droplet is assumed to be homogeneous according to the average density for different droplet sizes.~\cite{harms1998density,stringari1987systematics}
%The initial position of the photoelectron is randomly selected with requiring an exponential decrease in probability along the direction of the photon beam according to the absorption coefficient calculated from the atomic photoionization cross-section.~\cite{samson2002precision}
%In the experiment, the HNDs are formed in size-distributions. To account for this in the simulations, the simulation values and spectra are added up with there corresponding probability factor. The probability distribution varies in the three different expansion regimes.~\cite{toennies2004superfluid} In the subcritical expansion, helium atoms in the gas phase condensate to form droplets. The size distribution follows a log-normal distribution with a width comparable to the mean size.~\cite{harms1998density} In the supercritical expansion regime, liquid helium is expanded from which the droplets are formed, which leads to a exponential size distribution.~\cite{knuth1999average} In the critical expansion regime, a mixture of the two processes is present leading to combination of the two size distributions. The critical expansion temperature depends on the expansion pressure varying from 10-12.5~K for expansion pressures between 20 and 50 bar. 

\begin{figure}[t]
    \centering
    \includegraphics[width=0.95\columnwidth]{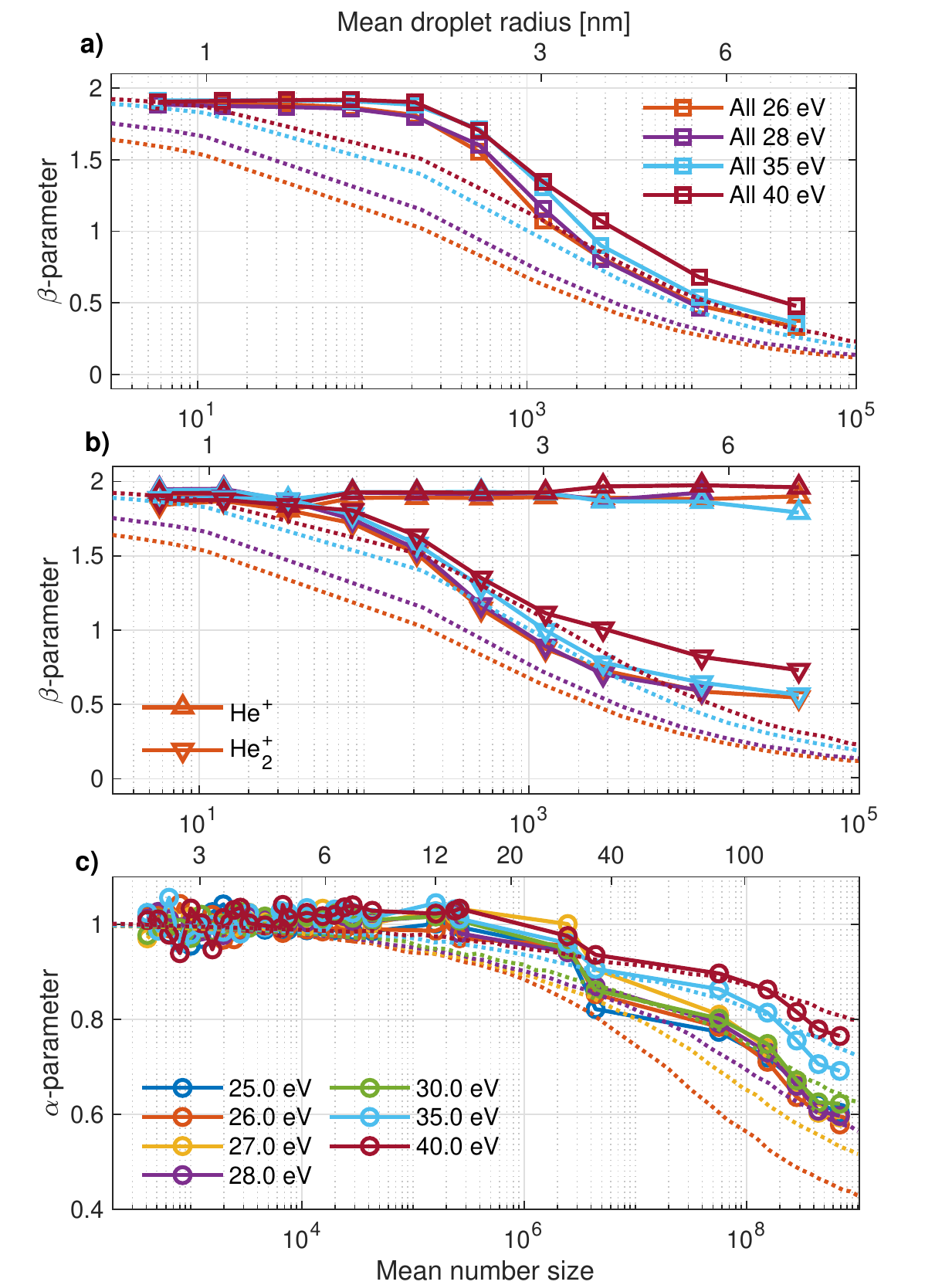}
    \caption{a) Anisotropy parameter $\beta$ inferred from VMIs as a function of mean droplet size at different photon energies for all electrons and b) for electrons detected in coincidence with He$^+$ and He$_2^+$. c) Shadowing parameter $\alpha$ as function of droplet size. Note the different $x$-axis range. The dotted lines show the corresponding results of the scattering simulations.
    %d) Ratio of trapped electrons to the estimated total number of photoionization events. The dotted lines in a)-d) show the corresponding results from classical scattering simulations. e) Experimental and simulated electron-energy loss spectra for two HND sizes recorded at $h\nu = 44$~eV. The colored dotted lines are best fits to the experimental and the simulated data. f) Mean relative energy loss as function of HND size determined from the fits of the electron spectra. Black triangles indicate values measured by Loginov \textit{et al.}~\cite{loginov2005photoelectron} 
    %*** I suggest to use the same bottom and top x-scales as in the other panels: Mean number size on the bottom, mean droplet size on the top. ***
    }
    \label{fig:graphs}
\end{figure}

Fig.~\ref{fig:VMIs} shows VMIs recorded at a photon energy $h\nu=26$~eV for various HND sizes. Photoelectrons emitted from free He atoms [Fig.~\ref{fig:VMIs}~a)] by absorption of one photon are emitted according to the PAD $I(\theta)\propto \cos^2(\theta)$ corresponding to the asymmetry parameter $\beta=2$, as conventionally defined by the angular intensity distribution $I(\theta) \propto 1 + \beta (3\cos^2\theta-1)/2$~\cite{cooper1968angular}. Here, $\theta$ is the angle between the light polarization and the emission direction. For increasing HND size, the angular distribution becomes more isotropic ($\beta\rightarrow 0$) due to the increasing number of elastic collisions the electrons undergo on their way out of the droplet [Fig.~\ref{fig:VMIs}~b) \& c)]. Fig.~\ref{fig:VMIs}~d) shows the VMI in the case where the HNDs are on average larger than the penetration depth of the incident light. At $h\nu=26$~eV, the photoionization cross-section of He is 6.79~Mb~\cite{samson2002precision} corresponding to a penetration depth of $67.6$~nm into liquid helium. 
For large HNDs, the PAD is dominated by the shadowing effect and Abel inversion is ill-defined due to breaking of the cylindrical symmetry about the polarization of the light meaning that we cannot define $\beta$ from the VMI. 
Additional VMIs recorded for different droplet sizes and photon energies are shown in Suppl. Fig.~1 in the Supplementary Information (SI)~\cite{SI}.

%The photoelectron angular distribution (PAD) is particularly sensitive to elastic electron scattering~\cite{ohrwall2003observation} and is often described by the asymmetry parameter $\beta$ defined by $	I(\theta) \propto 1 + \frac{\beta}{2}(3\cos^2\theta-1)$. 

% *** First describe all panels of this fig.; State the main features, at which sizes, and briefly state the obvious interpretations. This is an overview of the phenomenology and sets the stage for a more detailed discussion in the following.

Fig.~\ref{fig:graphs}~a) shows $\beta$ for all emitted electrons as a function of the HND size in the range $\langle N \rangle = 10$-$10^5$~atoms where the shadowing effect is absent. $\beta$ is constant up to $\langle N \rangle \sim 200$. Thereafter, it drops to 0.5 for $\langle N \rangle \sim10^4$. However, $\beta$ does not significantly depend on the photon energy, as previously observed~\cite{buchta2013extreme}. Fig.~\ref{fig:graphs}~b) shows $\beta$ as a function of HND size for electrons detected in coincidence with He$^+$ and He$_2^+$ ions, respectively. Electrons coincident with He$^+$ feature nearly constant anisotropy with $\beta\approx2$ for all droplet sizes; These electrons and He$^+$ ions stem from free He atoms accompanying the droplet beam even under expansion conditions (low nozzle temperature $\lesssim 12$~K) when large HNDs are formed. In contrast, for electrons detected in coincidence with He$_2^+$, $\beta$ decreases from 2 for $\langle N\rangle\lesssim30$ to 0.5 for $\langle N \rangle\approx 10^4$. As He$_2^+$ are formed by fragmentation of He clusters and HNDs, the electrons' PADs recorded in coincidence with He$_2^+$ are indicative for HND photoionization.
Note that for these electrons, $\beta$ drops below 2 for droplet sizes $\langle N\rangle >20$ whereas for all electrons [Fig.~\ref{fig:graphs}~a)], $\beta\approx2$ up to $\langle N\rangle\approx200$. This shows that in the regime of small droplets, $\langle N\rangle \lesssim 200$, free He atoms are more abundant in the supersonic jet than He droplets.
The electron mean-free path (EMFP) for elastic scattering calculated from the total electron scattering cross-section~\cite{adibzadeh2005elastic} and the number density of bulk liquid He~\cite{harms1998density} is 8~\AA~for an electron kinetic energy of 1~eV and it is 16~\AA~at 10~eV. This value is consistent with the He droplet radius $R\approx 1.2$~nm up to which $\beta\approx 2$ [Fig.~\ref{fig:graphs}~b)].

For larger He droplets, the $\alpha$-anisotropy due to the shadowing effect dominates the PADs. Fig.~\ref{fig:graphs}~c) shows that $\alpha$ decreases (increasing shadowing effect) as the droplets grow larger than $\langle N \rangle\approx 10^6$ ($R\approx 20$~nm).
%, which approximately matches the penetration depth of the light into the droplets. 
The size-dependence of $\alpha$ for electrons detected in coincidence with He$_2^+$ (correlated with HNDs) closely follows the one for all electrons because the abundance of free He atoms in the jet is small under these conditions (see Suppl. Fig.~3~\cite{SI}). 

\begin{figure}[t]
    \centering
    \includegraphics[width=0.95\columnwidth]{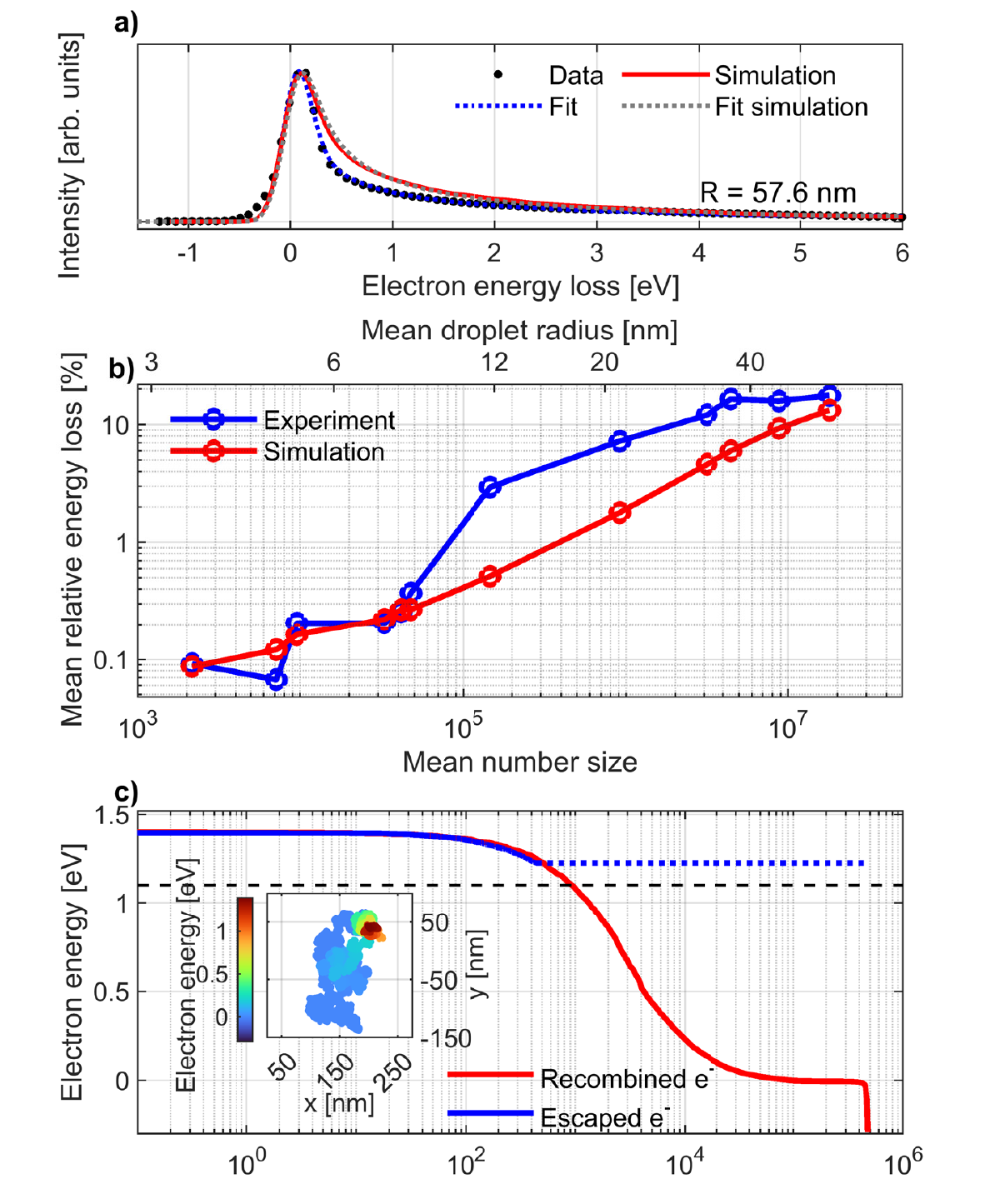}
    \caption{
    a) Experimental and simulated EEL spectra for large HNDs ($R=57.6~$nm) recorded at $h\nu = 44$~eV. The colored dotted lines are best fits to the experimental and simulated data. b) Mean relative energy loss as function of HND size determined from the fits of the electron spectra. 
    %Black triangles indicate values measured by Loginov \textit{et al.}~\cite{loginov2005photoelectron}
    c) Classical simulations of electron scattering in a HND at a photon energy of 26~eV. The inset shows an electron trajectory for the case where the electron loses all its energy and returns to its parent ion where it recombines. The electron energy is encoded in the line color.
    %The $x$ position is along the light propagation direction (in the $-x$-direction), and the y-axis the polarization axis of the incoming light.
    The evolution of the electron energy is shown as red line in the main panel. The blue line indicates a trajectory where the electron escapes from the droplet. The black dashed line shows the potential barrier for electron localization in the bulk of large HNDs.     %Trajectory for the electron escaping can be seen in the supporting information. 
    }
    \label{fig:traj}
\end{figure}

% A significant portion of the photoelectrons do not escape the droplet and are, therefore, not detected (see fig. \ref{fig:graphs} d). Instead the electron recombines with the He cation or is trapped in a bubble~\cite{gonzalez2020solvation}. The proportion of electrons trapped in HNDs is challenging to estimate since we do not have a direct measure of the total photoelectrons created in the experiment. We estimate the total number of photoionization events by assuming that all photoelectrons are detected in the case of small droplets ($\langle N \rangle < 10^5$) and that the total number of photoionization events is proportional to the target density in the HND beam. The relative increase in target density is determined from the stagnation pressure measured in the HND beam dump taking the beam velocity into account~\cite{gomez2011sizes}. The relative electron number loss starts increasing for ($\langle N \rangle > 10^5$) to having $\sim90$\% of the electrons recombined or trapped in a bubble for $\langle N \rangle > 10^8$. Since the relative electron loss is determined from a non-direct measure relying on assumptions, the final value should be treated with appropriate uncertainty. 
%The rise in electron loss for large HNDs shows that an increasing amount of the electrons lose nearly all of their kinetic energy from scattering in the droplet.   
%Since the cylindrical symmetry of the VMI about the polarization axis is broken for large HNDs due to shadowing, the electron energy cannot be determined from Abel inversion of the VMIs~\cite{dick2014inverting}. 
To investigate the energy loss of the photoelectrons for large HNDs, we recorded high-resolution PESs using a hemispherical analyzer. Fig.~\ref{fig:traj}~a) shows an EEL spectrum recorded at $h\nu = 44$~eV (see electron spectra for other droplet sizes in Suppl. Fig.~11~\cite{SI}). 
%Note that the PESs recorded with the hemispherical analyzer are recorded for droplet sizes where shadowing in not dominant yet since otherwise the electron intensity would depend on the mounting angle of the analyzer with respect to the incoming direction of the light.  
As the droplet size increases, the photoemission line develops a tail toward higher EEL (lower electron energy). 
%This trend is reproduced by the simulated electron spectra, which are convoluted with an instrument function taken as a Gaussian function (blue lines in Fig.~\ref{fig:graphs} e). 
Similar asymmetric line broadening was previously reported for small HNDs ($R=2$-$6$~nm) doped with aniline molecules~\cite{loginov2005photoelectron}. While the line shape was described by an exponentially decaying function $\propto\exp[-E_e/a]$, here we find empirically the best fit with the function $\propto\exp[-(E_e/a)^b]$ (see dotted lines in Fig.~\ref{fig:traj}~a) and details in the SI~\cite{SI}).
From these fits we infer the mean relative EEL (see Suppl. Fig.~11~\cite{SI}), which is shown in Fig.~\ref{fig:traj}~b) as function of HND size.
Note that the EEL only pertains to the electrons emitted from the droplets. The total EEL is in fact higher if all electrons trapped in the droplets are taken into account. For a droplet of $R\sim50$~nm, $\sim15$\% of the electron energy is lost on average due to elastic scattering. This corresponds to $\sim300$~binary collisions according to simple estimates of the kinematics assuming head-on collisions (see SI~\cite{SI}). 
% The relative EEL found from the experiment with aniline-doped HNDs shows a similar increase with droplet size; However, the relative EEL found here is somewhat smaller, possibly due to the different photoionization scheme.
   
To obtain a better understanding of the electron-He interaction leading to changes in the PAD, the experimental results are compared with classical electron–He elastic scattering simulations based on differential (energy, scattering angle) scattering cross sections. See the SI~\cite{SI} for a detailed description. 
% In the comparison between experiment, we neglect droplet size distribution in the experiment assuming $N = \langle N \rangle$~\cite{harms1998density,knuth1999average}. 
In the simulation, an electron recombines with its parent ion when it turns around and binds to the ion with a binding energy exceeding its kinetic energy. A simulated electron trajectory for this case is displayed in
Fig.~\ref{fig:traj}~c) as an inset. The corresponding time evolution of the kinetic energy is shown as a red line. For comparison, a trajectory where the electron escapes the droplet is included (blue line). 
%Thus, electrons can lose their entire kinetic energy within 10-100~ps by elastic scattering before recombining with their parent ions after several 100~ps and more than 100~nm excursion in the simulation. 
Large HNDs, which have similar properties as bulk superfluid He, feature a potential energy barrier of $\sim$1.1~eV below which electrons localize in bubbles [black dashed line in Fig.~\ref{fig:traj}~c]~\cite{broomall1976density,henne1998electron,wang2008photoelectron}. Thus, in the simulated PAD we exclude those LEEs with kinetic energy $<1.1~$eV. Obviously, this leads to an inconsistency between the simulation and the experiment at $h\nu < 25.7$~eV indicating that not all low-energy electrons are actually trapped, \textit{e.~g.} those formed near the surface. %This indicates a more complicated electron trapping dynamics that a simple barrier potential cannot account for.  
%However, the simulated electron spectra include the many electrons that lose most of their kinetic energy by elastic scattering, but never recombine with the ion within the simulation time. Thus, the simulation tends to overestimate the number of emitted LEEs compared with the experimental VMIs (see SI~\cite{SI}). In real HNDs, these LEEs are most likely solvated by forming bubbles and eventually desorb from the droplets or recombine with ions~\cite{farnik1998differences}. 
We do not have a direct measure of the number of photoelectrons evading detection due to trapping in the HNDs, but can only make a rough estimate based on the He flux measured by the pressure increase in the HND beam dump (see Suppl. Fig.~6~\cite{SI}).
%The simulated $\beta$-parameter matches with PAD for electrons detected in coincidence with He$_2^+$ which are the droplet-correlated ones. 

The results of the simulations including the 1.1~eV cutoff are shown as dotted lines in Fig.~\ref{fig:graphs}. For reference, the results of the simulation without energy cutoff are shown in the Suppl. Fig~10~\cite{SI}. The simulated $\beta$ values follow a similar size-dependence as those inferred from electrons detected in coincidence with He$_2^+$ in the experiment. 
%while $\beta$ from the measured total electron PAD reaches the atomic value for larger droplets than the droplet-correlated coincidence electrons. This is due to an increasing contribution to the detected electrons by uncondensed He atoms produced by the expansion at higher stagnation temperatures.
%The angular anisotropy associated with the photoionization process is already reduced for He droplets with sizes $\langle N \rangle > 30$ as a result of elastic electron-He scattering. 
The PAD becomes more isotropic as $\langle N \rangle > 30$ due to elastic electron-He scattering. Deviations from atomic PAD have been observed for inner-valence ionization of the clusters of heavier rare-gas atoms in a size range from $\langle N \rangle = 70$-$7000$~\cite{zhang2008angular,pfluger2011observation,ohrwall2003observation,rolles2007size}. The conclusion was that PAD are significantly changed when the cluster size reaches the magnitude of the EMFP for elastic scattering~\cite{zhang2008angular}, which is consistent with our findings for HNDs.
% *** Add a sentence about EMFP vs. elastic scattering simulations, which are based on angle- and energy dependent cross sections (it's not exactly clear to me where the total elastic scattering cross section comes from and how it deals with the large probability of nearly forward scattering at low energies?) ***
% In water clusters, the PAD is significantly changed already for small clusters consisting of $N\geq 20$ molecules~\cite{hartweg2017size}. *** only elastic or also inelastic scattering? ***
The shadowing effect for large HNDs is also well reproduced by the scattering simulations, see Fig.~\ref{fig:graphs}~c). The corresponding drop of $\alpha$ for $\langle N\rangle\gtrsim10^6$ is more pronounced near the He ionization threshold (24.59~eV) where both the absorption cross-section (7.40~Mb~\cite{samson2002precision}) and the electron-He scattering cross-section (600~Mb~\cite{adibzadeh2005elastic}) are highest. Note that at higher photon energies $>45~$eV additional features appear in the electron VMIs due to inelastic scattering of photoelectrons on He atoms in the droplets, see Suppl Fig.~2~\cite{SI}. Different $\alpha$ values for these features inform about the mechanisms of generating these electrons.

The value $\alpha=0.6$ measured for $R=100$~nm at $h\nu=25$~eV is close to that reported for solid salt NPs (NaCl~\cite{wilson2007size}, KCl~\cite{goldmann2015electron}) of similar size. 
%The fact that the shadowing effect is present in HNDs and that the effect can be described purely by elastic electron-atom scattering demonstrates that elastic scattering has a significant effect on energy loss of photoelectrons in condensed matter and possible recapturing of the electron.
% *** I don't agree: Alpha doesn't tell us about energy loss, does it? Does it tell us about electron recapturing? Needs to be explained ***
% The shadowing effect that we report here for HNDs must be caused by solvation of electrons in medium following significant energy loss from scattering. High kinetic energy electrons ($\gtrsim 10$~eV) are more favorable to scatter in the forwards direction which rules out shadowing induced purely from backscatter of the electrons~\cite{adibzadeh2005elastic}. 
%For heavier rare gas clusters, it was found that elastic scattering only contribute to creating more isotropic PAD~\cite{zhang2008angular}. 
%Shadowing being a result of electron energy loss and solvation following elastic scattering is confirmed by the fact that a majority of the created photoelectrons are not detected for large HNDs. 
%These LEEs are most likely solvated in the droplet through bubble formation, which can partly explain the large estimated electron number loss from the experiment which is not accounted for by electron-ion recombination in the simulation. The solvation dynamics of electrons in liquid helium as function of electron kinetic energy is not well understood. Therefore, we refrain from estimating the number of solvated electrons from the LEE distribtion in the simulation, but simply note that introducing a cut-off energy for electron solvation does not account the full electon number loss in the experiment. 
%%%%%%%%%%
The fact that shadowing becomes the dominant effect when the HND size approaches the photon penetration depth ($\sim 70$~nm) implies that the latter is the upper bound for the electron trapping range. A conservative estimate of the lower bound is the EMFP for elastic scattering ($\sim1$~nm). Thus, the trapping range is in order of 10~nm, which is consistent with the rise in electrons evading detection which we observe in the size range $R=10$-$40$~nm (see the Suppl. Fig~6~\cite{SI}). From the simulations we determine the trapping range by evaluating the maximum distance travelled by an electron from the cation in bulk liquid He, see Suppl. Fig.~7-9~\cite{SI}. For $h\nu<27$~eV, the trapping range is smaller than the photon penetration depth and it becomes similar in magnitude for $h\nu\gtrsim 27$~eV. 

When excluding the droplet barrier for electron trapping, the simulated values of the $\alpha$-parameter closely follow those found when the 1.1~eV-barrier is taken into account, but the trapping range is largely overestimated. As the the trapping range and the potential barrier are closely correlated, an alternative condition for LEE trapping in He droplets could be a finite escape depth at the droplet surface, see Suppl.~Fig.~9~\cite{SI}. The higher efficiency of electron-He elastic scattering observed in experiments as compared to the classical scattering simulations is likely due to quantum effects becoming important at low energies $\lesssim 10~$eV where the de Broglie wavelength of the electron reaches the average distance between He atoms in a HND (3.6~\AA) or even the size of the whole HND ($\lesssim 1$~eV). However, performing quantum-scattering simulations goes beyond the scope of this work.

The experimental and simulated electron spectra show a similar tail extending from the photoline toward higher EEL [see Fig.~\ref{fig:traj}~a)]. However, given that the resulting average relative EEL is only 15\,\% and the peak of the photoemission line remains nearly unshifted, we conclude that photoelectron spectroscopy is possible even for large droplets ($\langle N \rangle > 10^7$) provided the electron energy well exceeds 1.1~eV. 
The larger EEL found in the experiment with aniline-doped HNDs~\cite{loginov2005photoelectron} is likely due to the low energy of emitted electrons ($\sim0.8$~eV), at which scattering should be treated quantum mechanically and electron localization by formation of bubbles can no longer be neglected. 
In summary, we have demonstrated the effect of elastic scattering on the energy and angular distributions of photoelectrons emitted from HNDs in wide ranges of droplet size and electron energy. We identify two regimes of anisotropic electron emission. For small droplets ($\langle N \rangle < 10^6$), the cylindrical symmetry with respect to the polarization of the light is retained. However, electron emission becomes more isotropic compared to free He atoms when the HND radius $R\gtrsim1.2$~nm exceeds the EMFP for elastic scattering. Large droplets ($\langle N \rangle > 10^6$) exhibit a pronounced shadowing effect and the PAD becomes anisotropic along the light propagation direction. 
% The fact that shadowing takes place in HNDs demonstrates the strong effect of elastic scattering in photoionization studies of NPs and condensed phase medium not only on the loss of angular information, but also on the loss of kinetic energy of the photoelectrons and the possible solvation of them as consequence of this.
%*** does shadowing really indicate energy loss? ***. 
% The fact that both kinetic energy and angular distributions are strongly affected by elastic scattering is important to take into account in future experiments using HND as spectroscopic matrix for photoionization for dopant molecules or aggregates. 
% However, the detected photoelectrons lose only $15\,$\% of their kinetic energy on average for a droplet size of $\langle N \rangle \sim 10^7$ and for typical experimental conditions ($\langle N \rangle \sim 10^4$), the average energy loss is $<1$\%. 
For typical experimental conditions ($\langle N \rangle \sim 10^4$) used in spectroscopy experiments, the average EEL is only $<1$\,\%; for large droplets ($\langle N \rangle \sim 10^7$) the average EEL of the detected photoelectrons rises but stays below $15\,$\%. This makes HNDs a suitable matrix for UPS and XPS of embedded species, but a significant loss of electron angular information should be expected. Likewise, a large number of electrons remain undetected as multiple elastic scattering leads to trapping of the electrons in the droplets where they recombine with their parent ions.
% *** Better would be to say more specifically under which conditions it is still possible to measure PES of dopants; After all the distortion of the photoline isn't that bad, even for large droplets. Of course, there the electrons are only emitted out of the penetration region -- how deep is it? ***
% *** we should also add a sentence connecting back to radiation damage, in particular that induced by slow electrons; Could we even estimate how much less effective elastic scattering is for heavier atoms / molecules? ***
%%%% Rewrite
% Most of the photoelectrons are solvated in large HNDs ($\langle N \rangle > 10^6$), which shows that elastic scattering is an efficient channel to form LEEs in liquid He. To determine the importance of elastic scattering in forming genotoxic LEEs in biological matter, one needs to take into account the increased mass of the atoms in play. For carbon, nitrogen and oxygen, a 3-4 times smaller electron energy loss per elastic collision is expected. However, given the vast propagation of the electron in the scattering trajectory (more than 100~nm in HNDs), it could easily be expected that an electron would undergo sufficient elastic scattering events to form a LEE. 
%To determine the importance of elastic scattering in forming LEEs in molecular systems such as aqueous media, one needs to take into account the increased mass of the atoms in play. For carbon, nitrogen and oxygen, a 3-4 times lower electron energy loss per elastic collision is expected. 
In molecular systems such as water, electron scattering is affected by the higher mass of the molecular constituents and by their internal (ro-vibrational) degrees of freedom which open up additional inelastic scattering channels. Nevertheless, scattering of electrons on molecules where no internal modes are excited will contribute to EEL and should be taken into account as a mechanism generating highly reactive LEEs.

%Differentiating elastic and inelastic scattering in molecular systems is complicated by the possible excitation of intramolecular vibrations and intermolecular phononic modes. However our results demonstrate the significant effect one can get on electron energy loss from taken only elastic scattering. 

%However, given the vast propagation of the electron in the scattering trajectory (more than 100~nm in HNDs), it could easily be expected that an electron would undergo sufficient elastic scattering events to form a LEE in biological tissue.

\section{Acknowledgements}
J.D.A. and M.M. acknowledge financial support by the Carlsberg Foundation. We thank the Danish Agency for Science, Technology, and Innovation for funding the instrument center DanScatt. T.F. acknowledges support by the Deutsche Forschungsgemeinschaft (DFG, German Research Foundation) via SFB 1477 ``light–matter interactions at interfaces'' (project number 441234705) and via the Heisenberg program (project number 436382461). S.R.K. thanks Dept. of Science and Technology, Govt. of India, for support through the DST-DAAD scheme and Science and Eng. Research Board. S.R.K., K.S. and S.D. acknowledge the support of the Scheme for Promotion of Academic Research Collaboration, Min. of Edu., Govt. of India, and the Institute of Excellence programme at IIT-Madras via the Quantum Center for Diamond and Emergent Materials. S.R.K. gratefully acknowledges support of the Max Planck Society's Partner group programme. M.M. and S.R.K. gratefully acknowledge funding from the SPARC Programme, MHRD, India. 
A.R.A. acknowledges with gratitude for the support from the Marie Skłodowska-Curie Postdoctoral Fellowship project Photochem-RS-RP (Grant Agreement No. 101068805) provided by the European Union’s Horizon 2020 Research and Innovation Programme.
The research leading to this result has been supported by the project CALIPSOplus under grant agreement 730872 from the EU Framework Programme for Research and Innovation HORIZON 2020 and by the COST Action CA21101 ``Confined Molecular Systems: From a New Generation of Materials to the Stars (COSY)''.

%\nocite{*}

\bibliography{anisotropy_reference}% Produces the bibliography via BibTeX.

\end{document}

\title{Supplementary Information\\
Electron energy loss and angular asymmetry induced by elastic scattering in helium droplets}

%	\author{J. D. Asmussen \textit{et al.}}
%	\date{\today}

\author{Jakob D. Asmussen \textit{et al.}}
	
	\maketitle

	\renewcommand\refname{Supplementary References}
	\renewcommand{\figurename}{Supplementary Fig.}
	%\captionsetup[figure]{format=plain,labelformat=default}

\section{VMIs at different photon energies and HND sizes}
\begin{figure}[h]
    \centering
    \includegraphics[width=0.8\columnwidth]{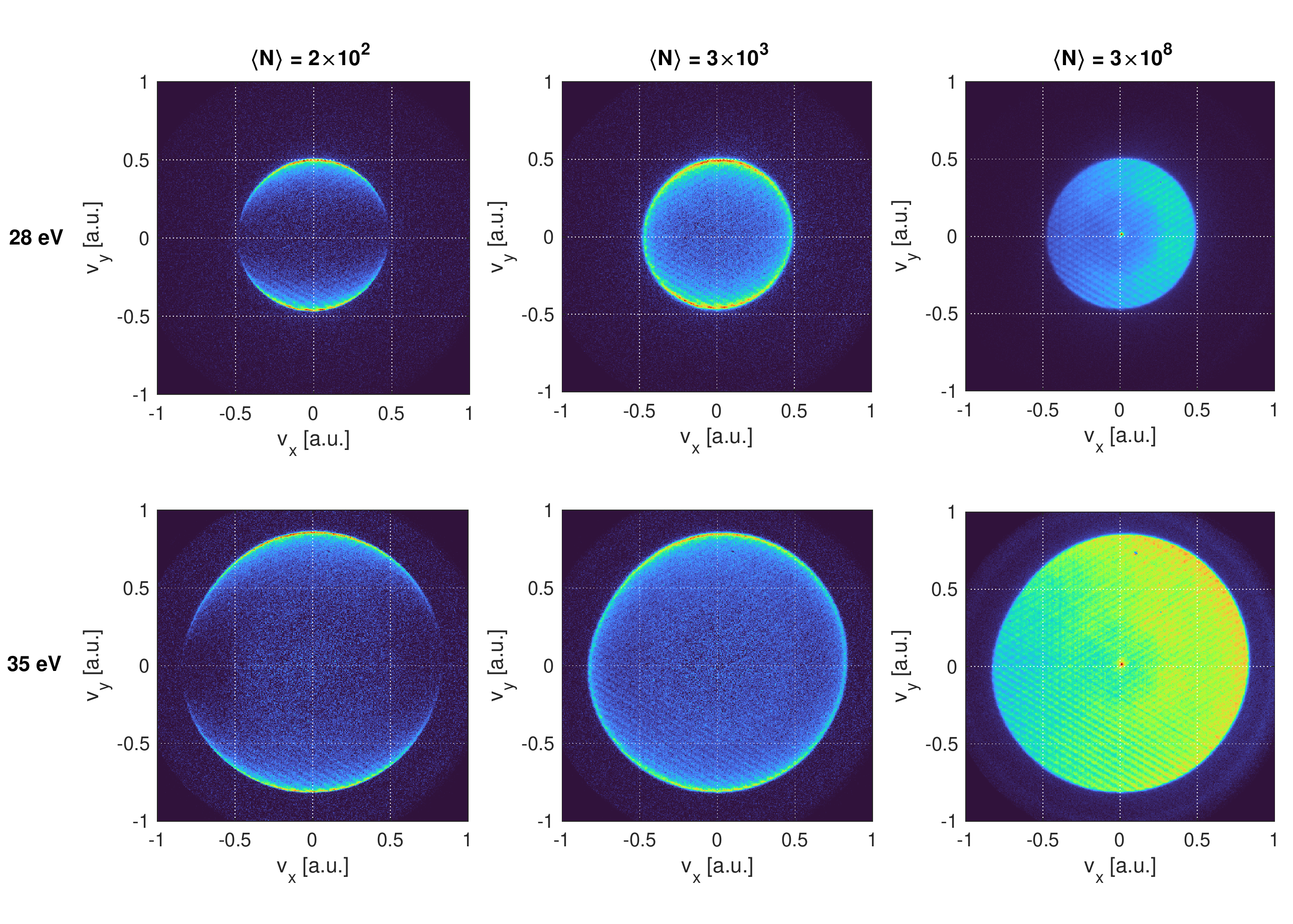}
    \caption{Experimental VMIs recorded for increasing He nanodroplet sizes (left to right) and increasing photon energy (top to bottom). }
    \label{SIfig:exp_VMI}
\end{figure}

Suppl. Fig.~\ref{SIfig:exp_VMI} shows VMIs recorded for three different He nanodroplet sizes and two photon energies ($h\nu=28~$eV in top row, $h\nu=35~$eV in bottom row). They illustrate three regimes of electron emission: Slightly perturbed PAD by elastic scattering (left column), almost fully isotropic PAD (middle column), and PAD dominated by shadowing (right column). 

\begin{figure}[h]
    \centering
    \includegraphics[width=0.5\columnwidth]{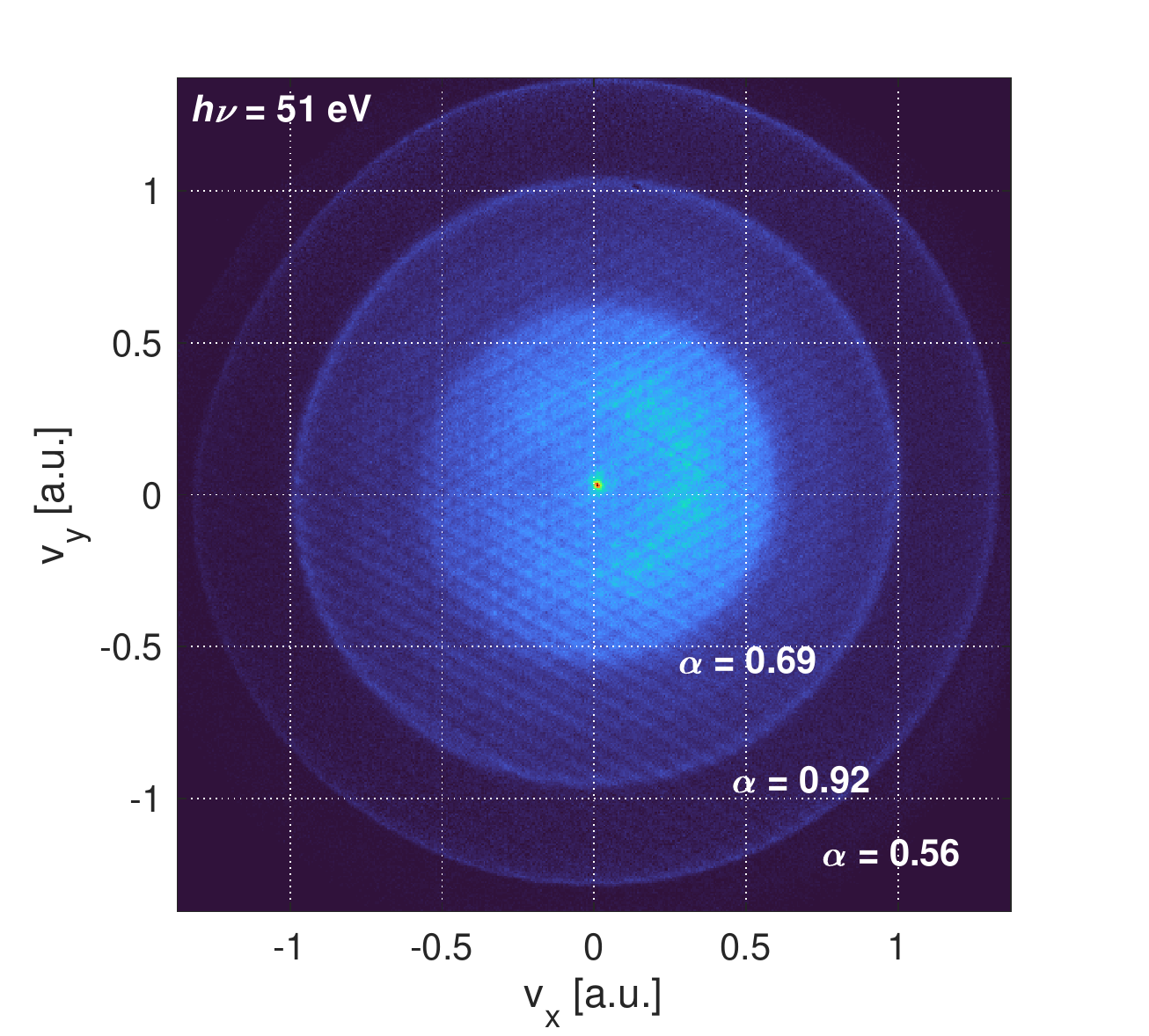}
    \caption{VMI recorded at $h\nu = 51$~eV for a droplet size of $\langle N \rangle = 2\times10^9$~atoms. The outer ring is the photoline ($h\nu - E_i$), the intermediate ring is due to electrons created by interatomic Coulombic decay of two excited He atoms and the inner circle is due to electrons inelastically scattered on another He atom in the same droplet before being emitted from the droplet. The $\alpha$-parameter for each contribution is given in the figure.  }
    \label{SIfig:VMI_inelastic}
\end{figure}

Suppl.~Fig.~\ref{SIfig:VMI_inelastic} shows a VMI recorded at $h\nu = 51$~eV for a droplet size of $\langle N \rangle = 2\times10^9$~atoms. At this photon energy, photoelectrons can inelastically scatter on He atoms in the droplets and excite or ionize another He atom in the same droplet~\cite{shcherbinin2019inelastic}. Three ring features appear in the VMI: The outer ring reflects the photoelectrons with energy $h\nu - E_i$, where $E_i=24.59~$eV is the ionization energy of He. The intermediate ring is due to interatomic Coulombic decay (ICD) of pairs of excited He atoms created by inelastic scattering~\cite{ltaief2023efficient}. The inner circle is due to inelastically scattered electrons. Both photoelectrons and inelastically scattered electrons show prominent forward/backward ($\alpha$) anisotropy due to the shadowing effect. The ICD electrons created by the secondary ICD reaction He$^*$+He$^*\rightarrow$He$^+$+He+$e_\mathrm{ICD}$ are emitted nearly isotropically ($\alpha=0.92$). This implies that the excited He$^*$ atoms, which are initially formed on the side of the droplet facing toward the incident light, roam around the He droplet and redistribute nearly isotropically over the droplet surface prior to the ICD reaction. Thus, this type of ICD is a slow process involving atomic motion on the length scale of the droplet circumference. 

\section{Anisotropic emission of coincidence electrons}
\begin{figure}[h]
    \centering
    \includegraphics[width=0.8\columnwidth]{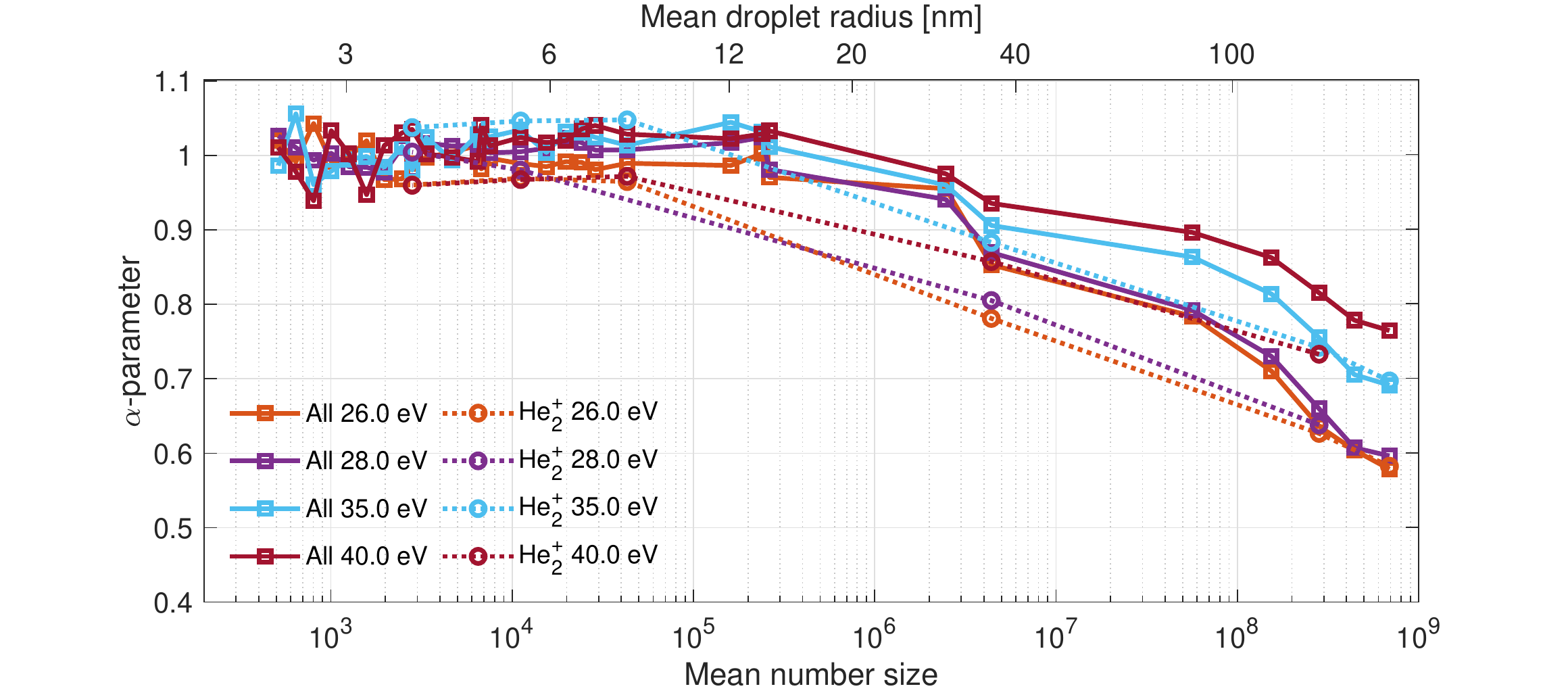}
    \caption{Anisotropy parameter $\alpha$ as a function of droplet size for all emitted electrons [reproduced from Fig.~2~c)] and for photoelectrons detected in coincidence with He$_2^+$.}
    \label{SIfig:alpha_He2}
\end{figure}

Suppl.~Fig.~\ref{SIfig:alpha_He2} shows the degree of shadowing characterized by the $\alpha$-parameter for all electrons and for electrons detected in coincidence with He$_2^+$. In contrast to small droplets where we analyzed the $\beta$ anisotropy [Figs.~2~a) and b)], the difference in $\alpha$ anisotropy between all electrons and electrons coincident with droplet-correlated cations is small since the atomic contribution in the beam is negligible when large droplets are formed in an expansion at low temperature ($T\leq 10$~K, expansion pressure is 30 bar). 

% *** Confusing: In Fig. 2 of the main text the dotted lines are simulations, here connecting lines between experimental data points. Should we add simulation curves? ***

\section{Electron scattering simulation}
\begin{figure}[h]
    \centering
    \includegraphics[width=0.8\columnwidth]{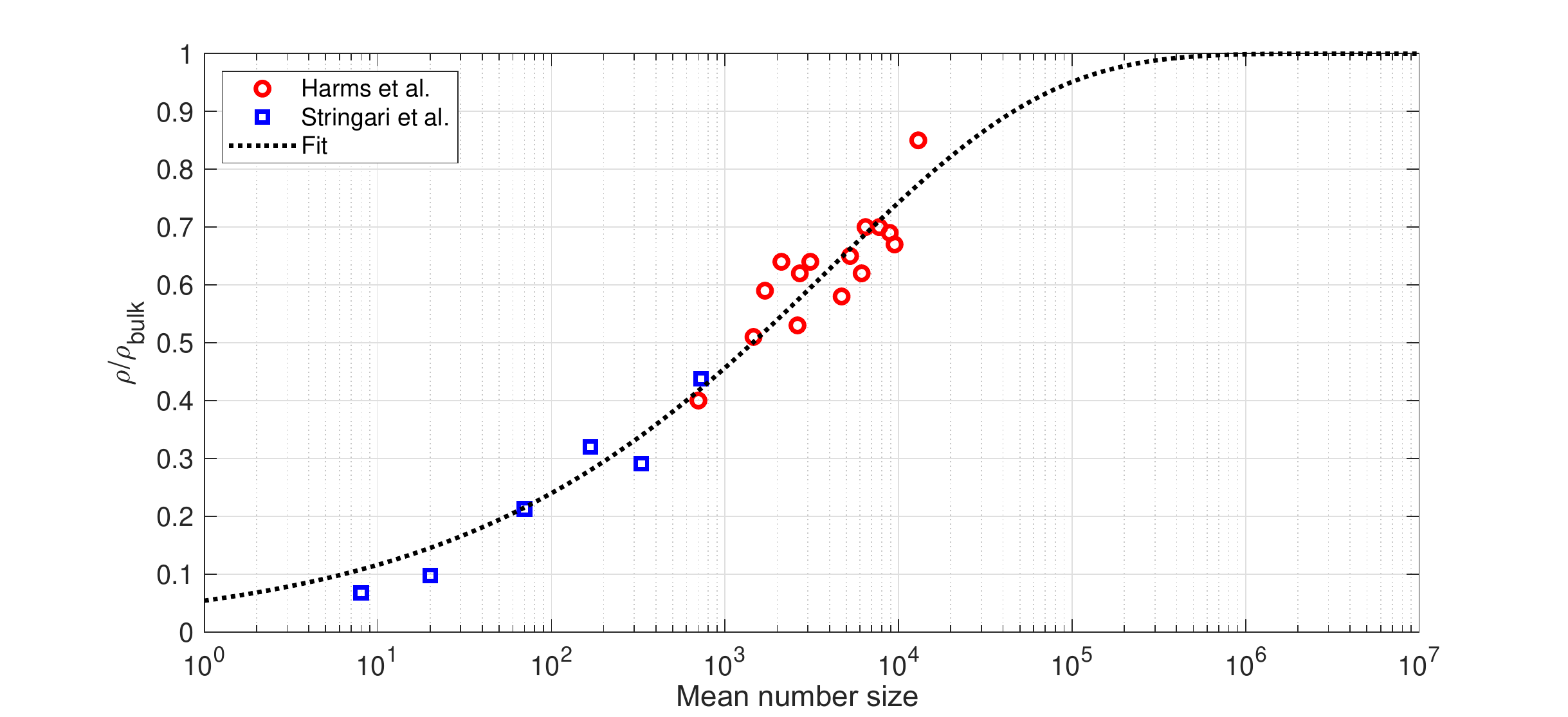}
    \caption{Mean He density relative to bulk density determined from fits of data reported by Harms~\textit{et al.}~\cite{harms1998density} and Stringari~\textit{et al.}~\cite{stringari1987systematics}.  }
    \label{SIfig:He_dens}
\end{figure}

The experimental results in this study are compared to an electron–He scattering Monte Carlo simulation based on doubly differential (energy, scattering angle) electron–He scattering cross sections~\cite{adibzadeh2005elastic} and the propagation of electrons along classical trajectories. In the simulation, the He number density inside the spherical droplet is assumed to be homogeneous. In reality, the droplet surface region has a variable density which crucially impacts the average density of small He clusters. We calculate the size-dependent average density in the droplet from the works of Harms~\textit{et al.}~\cite{harms1998density} and Stringari~\textit{et al.}~\cite{stringari1987systematics}. The average density ($\rho$) relative to bulk density of liquid He ($\rho_0$) as function of droplet number size ($N$) is found to follow the empirical formula 
\begin{equation}
    \frac{\rho}{\rho_0} = 1 - \exp{\left[-\left(\frac{N}{4155}\right)^{0.35}\right]},
\end{equation}
see the fit curve in Suppl.~Fig.~\ref{SIfig:He_dens}.

\begin{figure}[h]
    \centering
    \includegraphics[width=0.66\columnwidth]{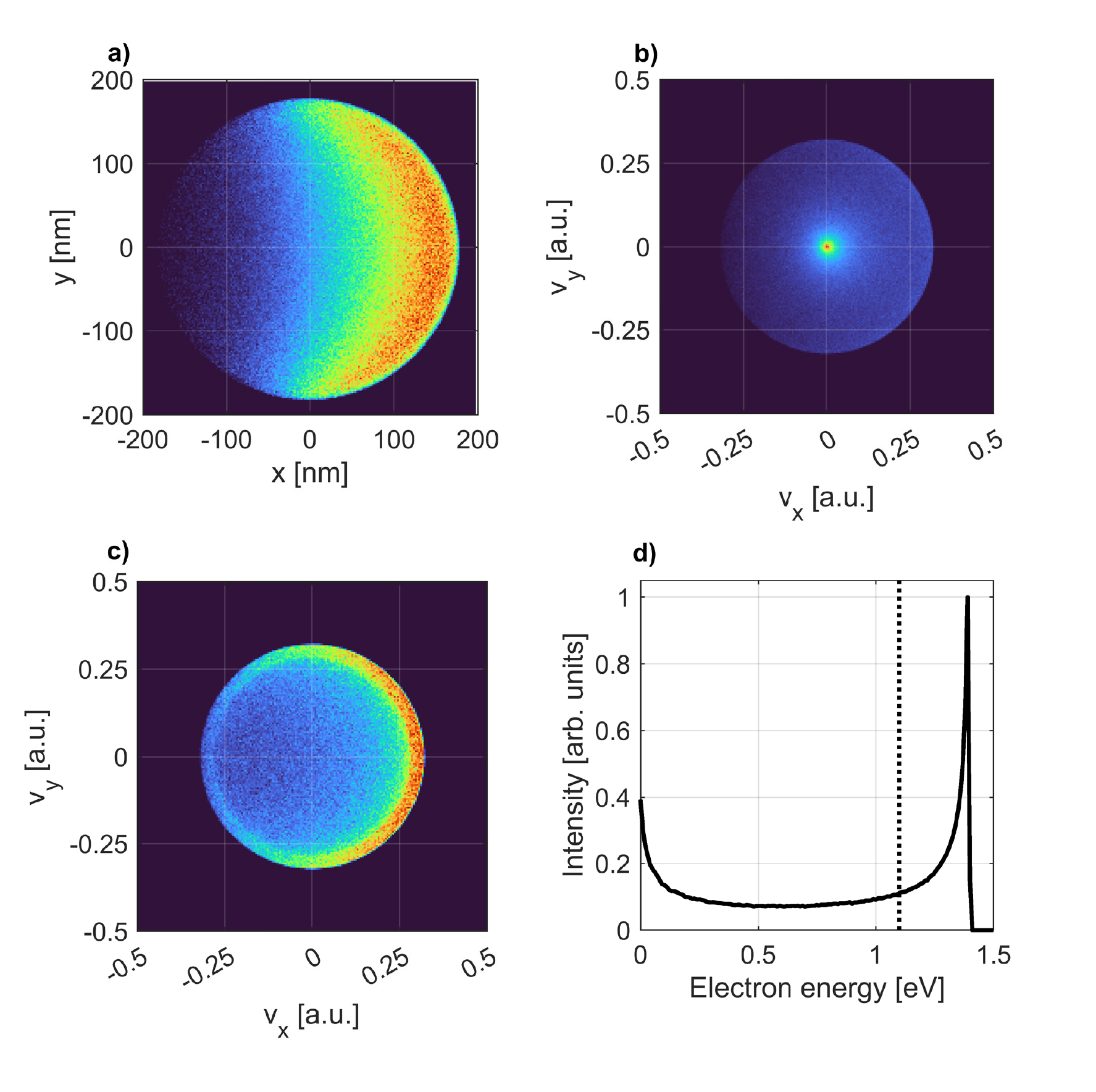}
    \caption{a) Cross-sectional view of the intensity distribution of the incident light onto a He droplet of radius $R = 170$~nm for at a photon energy of 26~eV. b) Projected velocity distribution of emitted electrons subjected to elastic scattering. c) Same projected distribution as in b) where electrons with kinetic energies $<1.1$~eV are excluded. d) Simulated electron spectrum corresponding to the same velocity distribution. The dotted line indicates the droplet potential barrier. }
    \label{SIfig:sim_VMI}
\end{figure}

% *** It's not exactly clear why this discussion? Is it that in the simulation, the this fit curve is used to define the droplet density for given size? This should be clearly stated. ***
In the comparison between simulation and experiment, we neglect the fact that in the experiment, droplet sizes are distributed according to a broad log-normal distribution, \textit{i.~e.} we assume $N = \langle N \rangle$~\cite{harms1998density,knuth1999average}. 
The initial positions of photoelectrons are randomly selected taking into account that the photoionization probability is exponentially attenuated inside the He droplet along the direction of the photon beam according to the absorption coefficient calculated from the atomic photoionization cross section~\cite{samson2002precision}. Suppl.~Fig.~\ref{SIfig:sim_VMI}~a) shows the absorption of the incoming light in a cross-sectional view of the droplet of radius $R=170$~nm. The photon energy of the incoming light is $h\nu=26$~eV, where the ionization cross-section of He is 6.79~Mb~\cite{samson2002precision} corresponding to a penetration depth of $67.6$~nm. %*** why only approximative value for the penetration depth when the cross section is quite precise? ***
The electron trajectory is propagated either up to a cut-off energy of 0 eV at which the electron is trapped in the Coulomb potential of the ion, or a cut-off energy 1.1~eV is included to account for solvation of the electron in the droplet which is assumed to occur at when the electron energy falls below the potential-energy barrier for electron localization~\cite{henne1998electron,broomall1976density}. 
Suppl.~Figs.~\ref{SIfig:sim_VMI}~b) and c) show the projected velocity distribution obtained from $10^6$ electron trajectories in the case of including and excluding the 1.1~eV potential barrier.
Excluding the barrier results in significant overestimation of the yield low-energy electrons (LEEs) in the electron spectra. The simulated VMI obtained from only electrons with energy higher than the potential barrier [Suppl. Fig.~\ref{SIfig:sim_VMI}~c)], the shadowing effect is clearly visible. Suppl.~Fig.~\ref{SIfig:sim_VMI}~d) shows the electron energy spectra corresponding to the VMI shown in Suppl.~Fig.~\ref{SIfig:sim_VMI}~b). The dotted line indicates the droplet potential barrier. The simulated electron spectra for large HNDs form a bimodal distribution. When the asymmetric photoline extends down to near-zero kinetic energy, LEEs accumulate at very low energy, migrating across the droplet in a diffusive motion over a large excursion distance [see Fig.~3~c)]. This behaviour is largely unrealistic; In the real system where low-energy electron-He scattering is governed by quantum-mechanical effects and nuclear motion sets in, electrons remain trapped in the droplets by localizing in bubble states.
% *** I would refrain from calling this a `state', sounds like quantum mechanics; Just say that the total electron energy becomes negative ***. 
%A significant part of the electrons escape the droplet with near-zero kinetic energy, which is not seen in the experiment. To visualize the shadowing effect in the PAD for electrons having non-zero kinetic energy, Suppl. Fig.~\ref{SIfig:sim_VMI}~c) shows the velocity distribution where electrons with $E_e<0.2$~eV are masked. The simulation overestimates the number of LEEs compared to the experimental VMIs [Suppl. Fig.~\ref{SIfig:sim_VMI}d)]. These LEEs are most likely solvated in the droplet by forming void bubbles. This likely explains the high loss of electrons in the experiment which is not accounted for by electron-ion recombination in the simulation.     

% *** So in the simulation the ZEKE's are drastically overestimated? Should be clearly stated / discussed. Why the masking of the ZEKE's? Is this the analysis procedure used to determine alpha? ***

\FloatBarrier

\section{Estimation of the fraction of solvated electrons}
\begin{figure}[h]
    \centering
    \includegraphics[width=0.8\columnwidth]{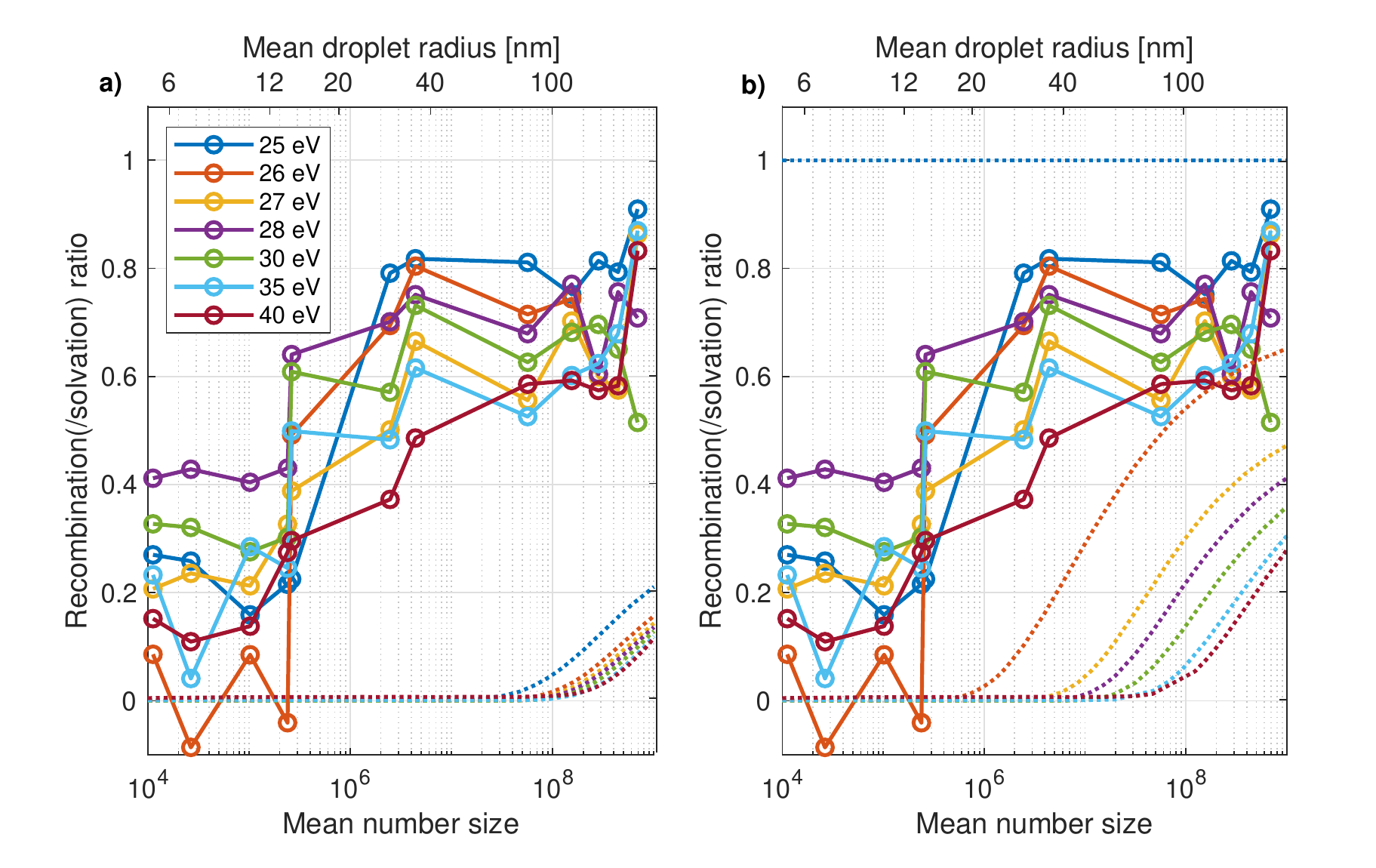}
    \caption{Ratio of trapped electrons to the estimated total number of photoionization events. The dotted lines show the ratio of electrons trapped in the droplets obtained from the simulation in the case of a) only including electron-ion recombination and b) introducing an energy cut-off of 1.1~eV to account for trapping of low-energy electron in the droplets. }
    \label{SIfig:solvation}
\end{figure}

%As seen in Suppl. Fig.~\ref{SIfig:sim_VMI}, LEEs are created by elastic scattering in the simulation. These electrons are not detected in the experiment but instead most likely solvated in the droplet. Additionally, LEEs recombine with their parent cation.  
The proportion of electrons trapped in HNDs is hard to estimate from the experiment since we do not have a direct measure of the total photoelectrons. We estimate the total number of photoionization events by assuming that all photoelectrons are detected in the case of small droplets ($\langle N \rangle < 2\times10^3$) and that the total number of photoionization events is proportional to the target density in the HND beam. The relative increase in target density is determined from the stagnation pressure measured in the HND beam dump taking the beam velocity into account~\cite{gomez2011sizes}. The shadowing effect is taken into account by integrating over the exponentially decaying photon intensity across the droplet. Suppl. Fig.~\ref{SIfig:solvation} shows the estimated proportion of solvated (and recombined) electrons from the experimental data. The relative electron number loss starts increasing for $\langle N \rangle > 10^5$ up to $\sim70$\,\% of the electrons being trapped for $\langle N \rangle > 10^8$. Since the relative electron loss is determined indirectly relying on the forementioned assumptions, the final values have large systematic uncertainties.
In Suppl.~Fig.~\ref{SIfig:solvation}~a), the proportion of trapped electrons from the simulation is shown for the case that the 1.1~eV potential barrier is not taken into account and trajectories are propagated up to electron-ion recombination occurs (dashed lines). Suppl.~Fig.~\ref{SIfig:solvation}~b) shows the electron trapping in the simulation including the potential barrier. Clearly, the number of trapped electrons is grossly underestimated when the barrier is excluded. 
%The classical scattering simulation includes only electron-ion recombination by setting the electron to be recombined as it reaches negative energy in the coulomb potential of the ion. 
%The solvation dynamics of electrons in liquid He as function of electron kinetic energy is not fully understood. 
Including the potential barrier results in an electron number loss $\sim60$\% at $h\nu=26$~eV for $\langle N \rangle > 10^8$. Evidently the model fails for $h\nu\leq 25.7$~eV where all electrons are trapped below the potential barrier in the simulation. In real He droplets, the He density drops to zero near the droplet surface; Clearly, a discrete 1.1~eV cut-off energy to account for electron trapping is a crude approximation. An improved model should account for the variable He density in droplets ranging from the bulk to very dilute He at the outer surface of the droplets. Additionally, quantum scattering and nuclear motion should be included.

\section{Electron trapping range}
\begin{figure}[h]
    \centering
    \includegraphics[width=0.6\columnwidth]{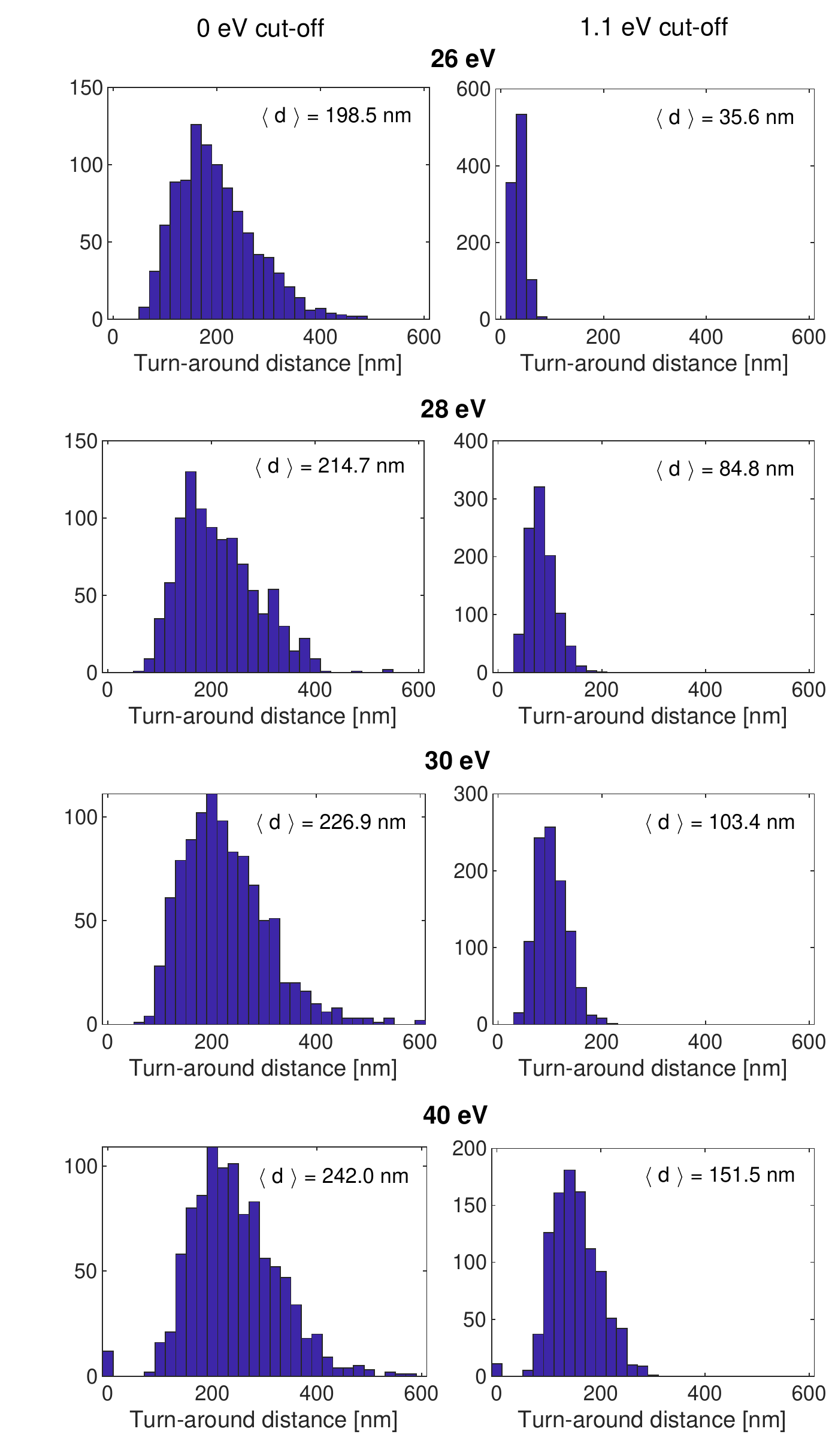}
    \caption{Maximum distance travelled by the photoelectron away from the cation (`trapping range') at different photon energies. The histograms in the left column correspond to trajectories propagated until the electron energy is $\leq 0$ eV (electron-ion recombination). The right column corresponds to trajectories propagated until the electron energies drop below the cut-off energy $\leq 1.1$~eV accounting for solvation of the electron in the He droplet. }
    \label{SIfig:turnaround}
\end{figure}

As stated in the main text, the photon penetration depth is an upper bound for the mean electron trapping range since shadowing becomes dominant when the droplet size approaches the photon penetration depth. To estimate this quantity more accurately, we determine the furthest distance that the electrons move from the cation in the simulation before turning around. Suppl.~Fig.~\ref{SIfig:turnaround} shows histograms of the maximum distance between the electron and cation in the trajectory for different photon energies. We show these histograms for two conditions of electron trapping -- excluding and including the droplet barrier potential. 
% Either the trajectory is ended at a cut-off energy of 0 eV meaning that only electron-ion recombination is included as electron trapping mechanism, or a cut-off energy 1.1 eV is included taking into account solvation of electrons with kinetic energy below the barrier potential of the droplet~\cite{henne1998electron,broomall1976density}. 
The average maximum electron-ion distance, $\langle d \rangle$, which corresponds to the mean distance the electron travels before being trapped, increases for increasing photon energy. By introducing the cut-off energy (potential barrier) of 1.1 eV, the average turn-around distance significantly drops. 

\begin{figure}[h]
    \centering
    \includegraphics[width=0.6\columnwidth]{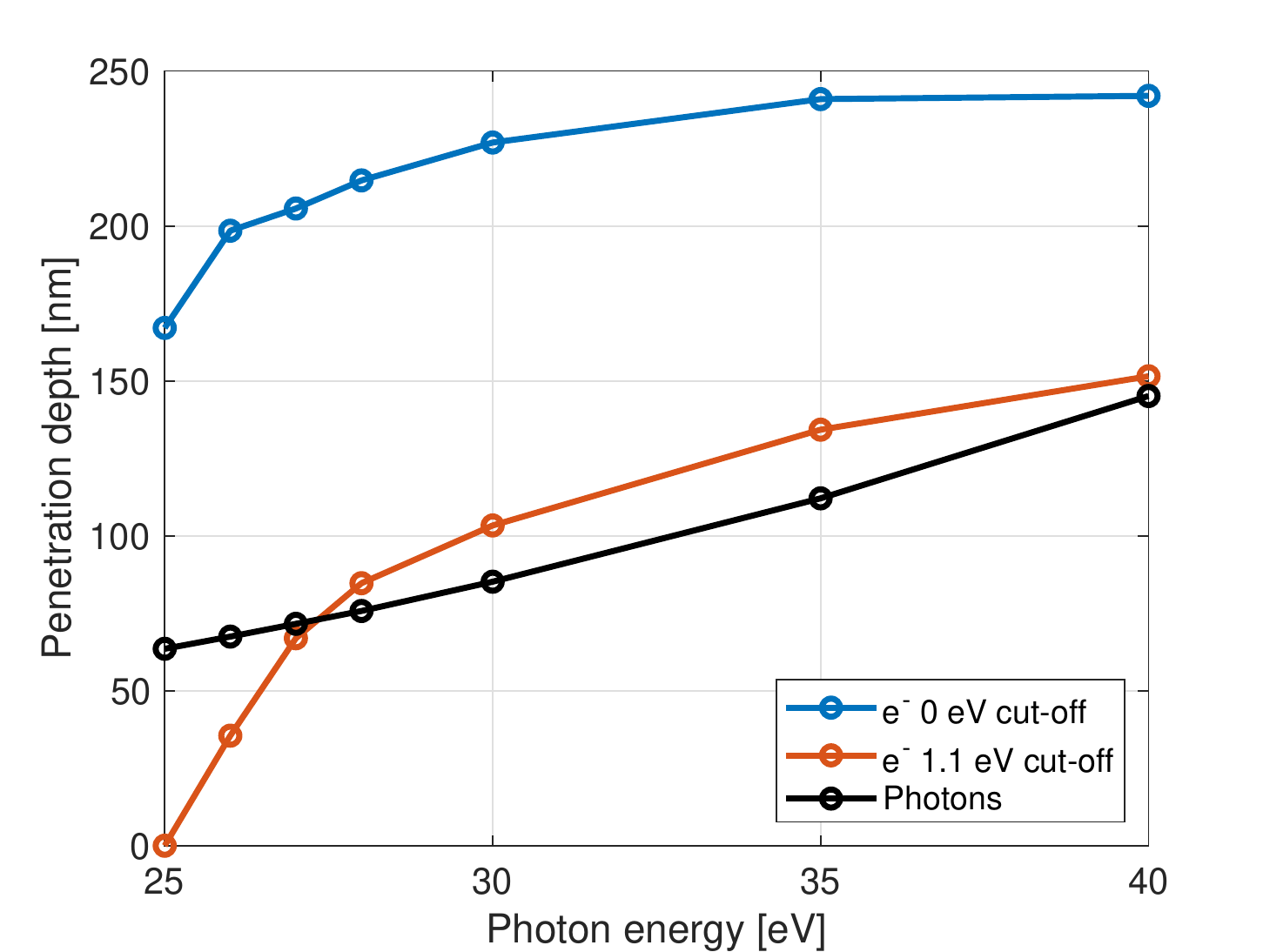}
    \caption{Photoelectron trapping range (blue and orange) and photon penetration depth (black) in HNDs as function of photon energy. The electron trapping range is determined from the average maximum distance that electrons travel away from their parent ions for two different electron cut-off energies.}
    \label{SIfig:penetration}
\end{figure}

Suppl.~Fig.~\ref{SIfig:penetration} shows the maximum excursion distance of the electrons from their parent ions (trapping range) as a function of photon energy for the two different cut-off criteria. For comparison, the photon penetration depth is shown, too~\cite{samson2002precision}. At $h\nu = 25$~eV, the cut-off energy of 1.1~eV cannot describe the real system since all electrons are formed with less kinetic energy than the barrier potential. However, as it can clearly be seen, photoelectrons from the droplets are indeed detected in the experiment. For $h\nu < 27$~eV, the electron penetration depth (with 1.1 eV cut-off) is smaller than the photon penetration depth, and for $h\nu\geq 27$~eV the electron trapping range and photon penetration depth are comparable in magnitude. When the droplet potential barrier is disregarded, the electron trapping range is much larger than the photon penetration depth which is inconsistent with the observation of shadowing in the experimental PAD. Therefore, the potential barrier is required to describe the electron dynamics in large HNDs.   
\FloatBarrier

\begin{figure}[h]
    \centering
    \includegraphics[width=0.7\columnwidth]{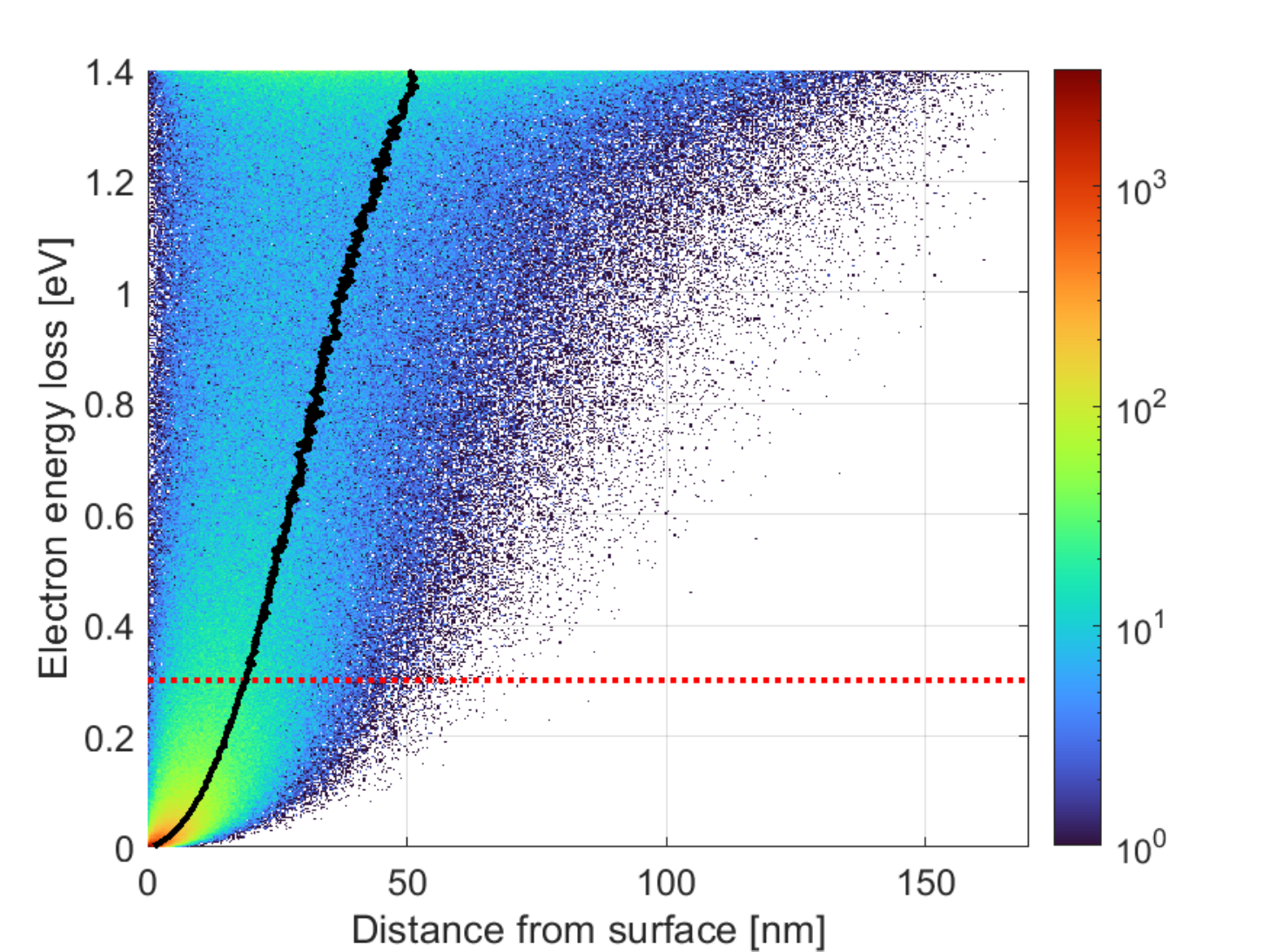}
    \caption{Density plot of the simulated energy loss of the escaped electrons as a function of the distance to the surface of droplet where the electron is created. The simulation is done for $h\nu = 26$~eV and $R=170$~nm. The red dotted line shows the droplet potential barrier. The black line shows the mean distance from the surface where an electron is formed to lose a given amount of kinetic energy ($y$-axis). }
    \label{SIfig:escape_depth}
\end{figure}

As a consequence of a limited electron mean free trapping range, only electrons formed closer to the droplet surface are more likely to escape the droplet with significant energy loss. Suppl.~Fig.~\ref{SIfig:escape_depth} is a density plot showing the initial distance of the photoelectron from the droplet surface as a function of simulated energy loss elastic scattering for $h\nu = 26$~eV and a droplet radius $R=170$~nm. The average electron energy loss becomes increasingly larger the deeper into the bulk the electron is created. Implementing the potential barrier of HNDs (red dotted line) means that only electrons formed $\sim50$~nm from the surface may escape the droplet. Due to the shadowing effect more electrons are formed near the surface of the droplet. Suppl.~Fig.~\ref{SIfig:escape_depth} also shows that electrons formed near the droplet surface either come out mostly with low energy loss or after losing most of their kinetic energy. This is because the initial electron velocity can point towards the surface or into the HND bulk where massive scattering occurs, respectively.
The resulting energy distribution has a bimodal structure, see Suppl.~Fig.~\ref{SIfig:sim_VMI} d).

The black line in Suppl.~Fig.~\ref{SIfig:escape_depth} shows the mean distance from the surface where an electron is formed to lose a certain amount of its kinetic energy. Thus, on average, electrons are formed $\lesssim19$~nm from the droplet surface to be able to escape the droplet when assuming a 1.1 eV cut-off energy. This matches well the turn-around distance shown in Suppl.~Fig.~\ref{SIfig:turnaround} for 26~eV. Thus, an electron with an initial kinetic energy of 1.4~eV travels a mean distance of $\sim20$~nm before being trapped in liquid He. The close correlation of the 1.1 eV cut-off energy to model trapping of LEEs and the finite electron escape depth suggests that the latter could be used as an alternative condition for electron trapping. However, both the energy cut-off and an escape depth are crude model assumptions; Instead, more sophisticated models of the interaction of LEEs with superfluid He should be developed.

\begin{figure}[h]
    \centering
    \includegraphics[width=0.6\columnwidth]{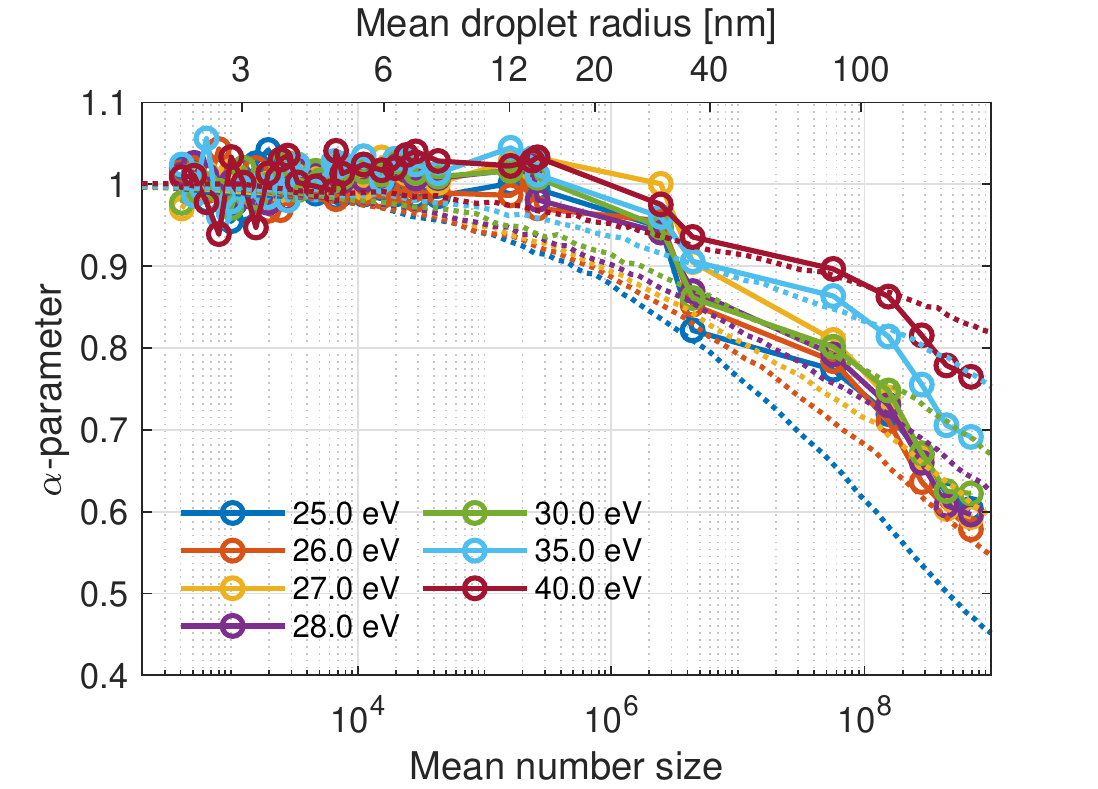}
    \caption{Shadowing parameter $\alpha$ as a function of droplet size. The dotted lines show the corresponding results of the scattering simulation for the case that the droplet potential barrier is not taken into account.}
    \label{SIfig:alpha_nocut}
\end{figure}

The $\alpha$-parameter determined in the simulation for the case that the droplet potential barrier is not taken into account matches nearly equally well the experiment as $\alpha$ calculated including the barrier (see Suppl.~Fig.~\ref{SIfig:alpha_nocut}). However, given the unphysically long excursion distance of the electron from its parent ion when no barrier is assumed, the values of the $\alpha$-parameter presented in the main text are inferred from simulations including the potential barrier. 

\FloatBarrier

\section{Fitting of electron spectra}
\begin{figure}[h]
    \centering
    \includegraphics[width=0.95\columnwidth]{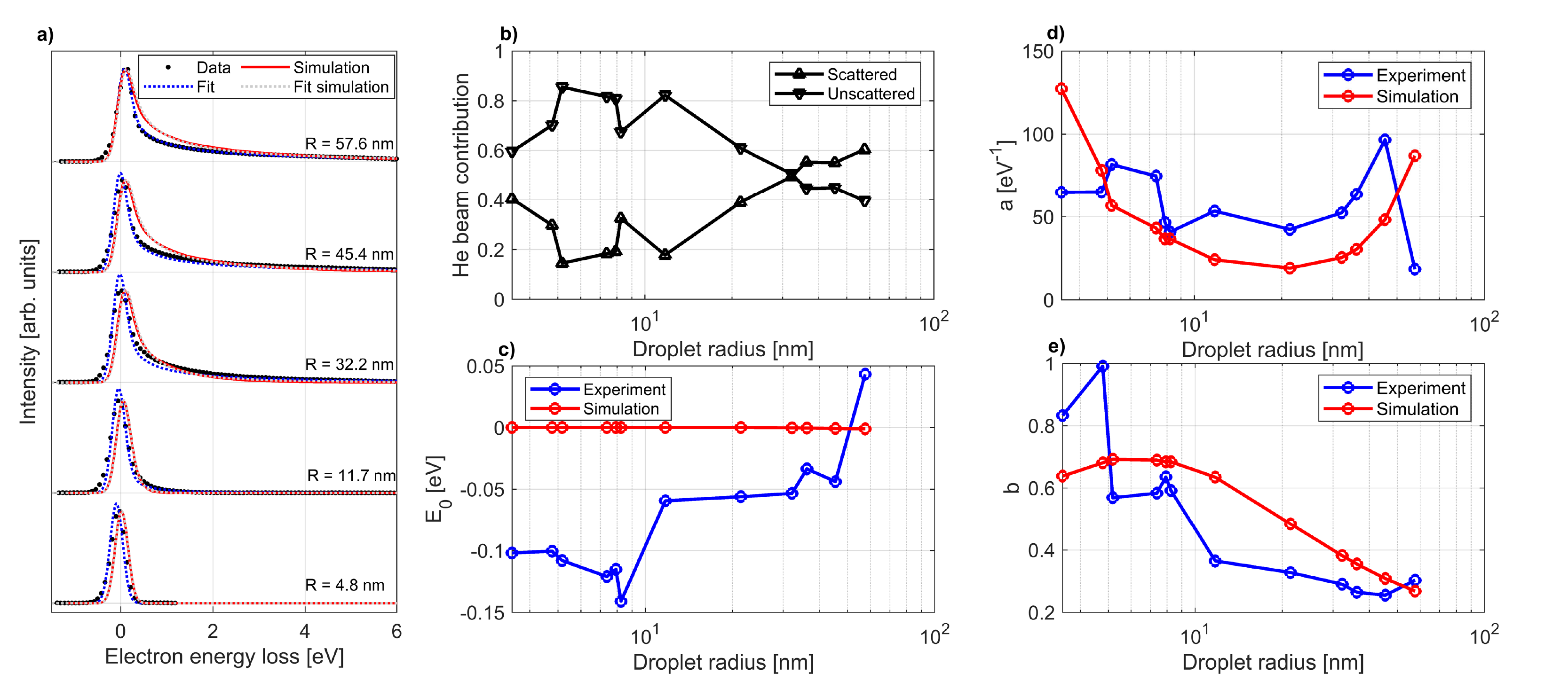}
    \caption{a) Measured and simulated electron spectra at $h\nu = 44$~eV for different droplet sizes and their corresponding fits. The simulated spectra are convoluted with a Gaussian function to take into account the experimental resolution which is determined from fits of the smallest-droplet spectra ($R=4.8$~nm). b-e) Values of the fit parameters as function of droplet size. In b), only the fit parameters to the experimental data are shown because only the fit to the experimental data includes the two different contributions (scattered and unscattered).   }
    \label{SIfig:elettra_fit}
\end{figure}

Electron spectra are recorded using a hemispherical analyzer at a photon energy of 44~eV. Suppl.~Fig.~\ref{SIfig:elettra_fit}~a) shows the electron energy loss for different droplet sizes. The spectra can be compared with simulated spectra which have been convoluted with a Gaussian function to account for experimental resolution. The width of the Gaussian is determined from fitting the spectra recorded for a droplet of size $R = 4.8$~nm. The electron energy loss tail is fitted using the following formula
\begin{equation}
f(E) = A \exp{\left[-\left( a (E-E_0)\right)^b\right]},    
\end{equation}
where $a$ and $b$ are free fit parameters with the limitation that $b$ is required to be $\leq1$. The fit function is numerically convolved with a Gaussian function similarly to the simulated spectra. For the experimental curve, the fit function is augmented with an additional Gaussian function taking into account an unscattered atomic part of the droplet beam. Suppl.~Fig.~\ref{SIfig:elettra_fit}~b) shows the contribution of the scattered and unscattered part of the beam from the fit. $E_0$ is introduced into the function to take into account any energy shift as function of droplet size and systematic errors in the detector calibration. An energy shift on the order of $0.1$~eV is found when increasing the droplet radius by one order of magnitude [Suppl. Fig.~\ref{SIfig:elettra_fit}~c)]. The absolute position of the peak may be affected by inaccuracies in the calibration. The fit parameters defining the extension of the energy loss tail ($a$ and $b$) are not found to converge to a single unique set of values. However, the average energy loss calculated from the two parameters are stable for different starting conditions of the fit. The $b$-parameter is larger (closer to 1) for smaller droplets matching with previous reports in the literature where the energy loss was found to be exponential for small droplets~\cite{loginov2005photoelectron}.

\begin{figure}[h]
    \centering
    \includegraphics[width=0.95\columnwidth]{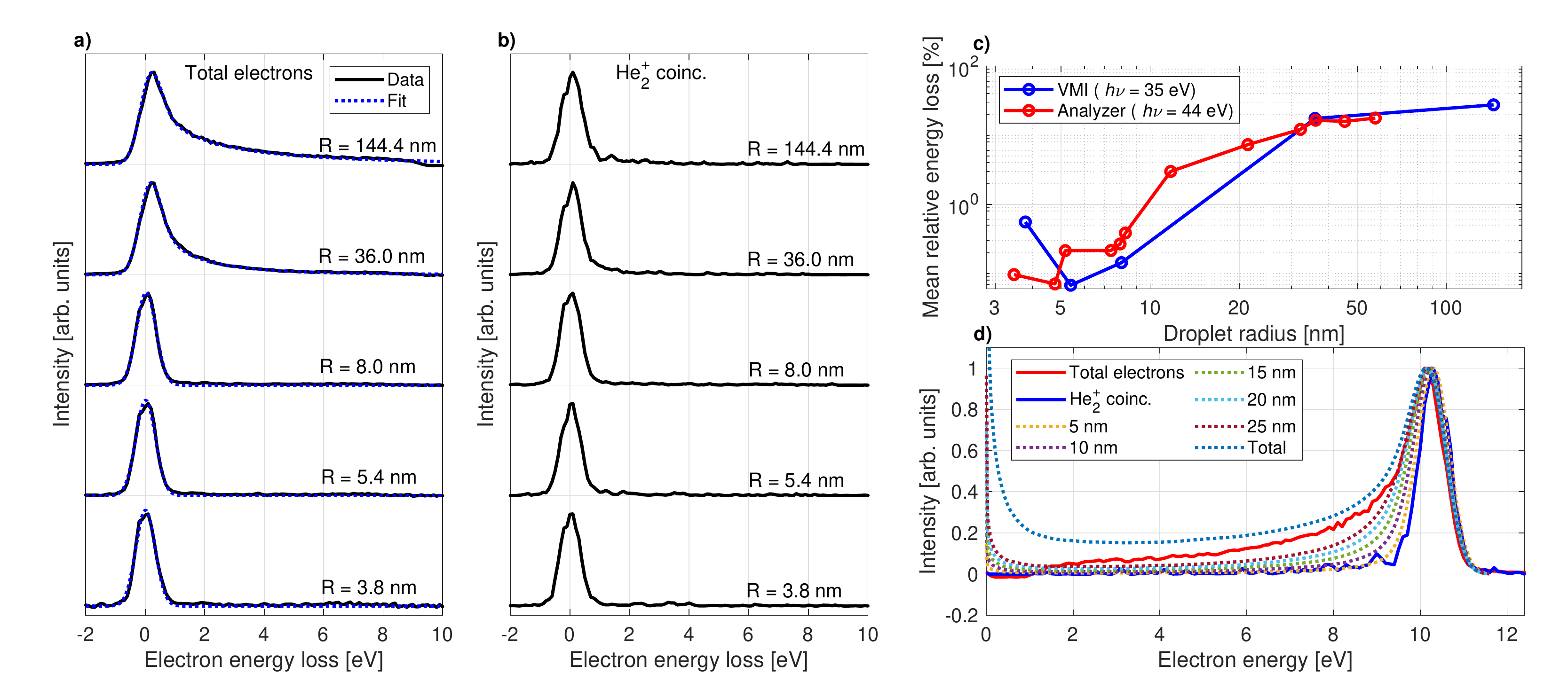}
    \caption{a-b) Electron energy loss spectra inferred from Abel inversion of VMIs recorded at a photon energy of 35~eV for total electrons and for electrons detected in coincidence with He$_2^+$. The spectra for all electrons are fitted using the outlined fit model, and the resulting mean relative energy loss is shown in c) in comparison with the size-dependent energy loss determined from the experiment using a hemispherical analyzer. d) shows the electron spectra inferred from VMIs for all electrons and for electrons detected in coincidence with He$_2^+$ for a droplet radius of $144.4$~nm. The dotted lines show corresponding simulations taking all electrons and only electrons formed in a surface layer of thickness $5$-$25$~nm into account. }
    \label{SIfig:astrid_fit}
\end{figure}

To validate the results from the fit model, we compare spectra recorded using a hemispherical analyzer with the velocity map images (VMIs) recorded at different droplet sizes. Due to the shadowing effect, the cylindrical symmetry of the VMIs is broken for large droplets and the Abel inversion of the images is not defined~\cite{dick2014inverting}. To circumvent this issue, we use only the backward plane of the image for Abel inversion. Suppl.~Fig.~\ref{SIfig:astrid_fit}~a) and b) show electron loss spectra from inversion of VMIs recorded at 35~eV photon energy of all electrons and electrons in coincidence with He$_2^+$-ions, respectively. The model fits the total-electron spectra well resulting in a mean relative energy loss matching the results from the spectra recorded with the hemispherical analyzer (see Suppl.~Fig.~\ref{SIfig:astrid_fit}~c). The electron spectra recorded in coincidence with He$_2^+$ ions show nearly no energy loss due to elastic scattering. By comparing with simulated electron spectra taking only electrons formed in a defined surface layer into account, Suppl.~Fig.~\ref{SIfig:astrid_fit}d) shows that only electrons formed in a surface layer of thickness of 5~nm contribute to the coincidences with He$_2^+$-ions. This demonstrates that He cations formed deeper inside the liquid bulk of the droplets are more likely solvated due to snowball formation~\cite{atkins1959ions}.      

\FloatBarrier
\section{Some estimates of the kinematics of electron-helium scattering}
The maximum energy transfer in a binary elastic collision of an electron with He atoms occurs in a head-on collision. Then, the energy loss per collision is~\cite{henne1998electron}
\begin{equation}
    \Delta E = E - E' = \frac{1}{2}mv^2-\frac{1}{2}mv'^2 = \frac{1}{2}mv^2 \left[ \left(\frac{M-m}{M+m}\right)^2 - 1 \right] \approx 4E\frac{m}{M}.
\end{equation}
Here, $v$ and $v'$ is the electron velocity before and after the collision, respectively, and $m = m_e$, $M = m_{He} \approx 8000\times m_e$. Thus, for $E = 10$~eV, this yields an energy
loss per collision of up to $\Delta E = 5$~meV.

The corresponding factor by which the electron energy is reduced is
\begin{equation}
    x = \frac{E'}{E} = \frac{v'^2}{v^2} = \left(\frac{M-m}{M+m}\right)^2 \approx 1 - 4\frac{m}{M} = 99.95\%.
\end{equation}
Thus, it takes $\approx300$ binary head-on collisions for a 15\% reduction of the electron kinetic energy. 

\bibliography{anisotropy_reference.bib}